\documentclass[sigconf]{acmart}

\AtBeginDocument{%
  }

\copyrightyear{2022} 
\acmYear{2022} 
\setcopyright{rightsretained} 
\acmConference[EAAMO '22]{Equity and Access in Algorithms, Mechanisms, and Optimization}{October 6--9, 2022}{Arlington, VA, USA}
\acmBooktitle{Equity and Access in Algorithms, Mechanisms, and Optimization (EAAMO '22), October 6--9, 2022, Arlington, VA, USA}
\acmDOI{10.1145/3551624.3555288}
\acmISBN{978-1-4503-9477-2/22/10}

\begin{document}

\title[Bias, Consistency, and Partisanship in U.S. Asylum Cases]{Bias, Consistency, and Partisanship in U.S. Asylum Cases: A Machine Learning Analysis of Extraneous Factors in Immigration Court Decisions}

%

\author{Vyoma Raman}
\authornote{Both authors contributed equally to this research.}
\email{vyoma.raman@berkeley.edu}
\affiliation{%
  \institution{University of California, Berkeley}
  \streetaddress{2224 Piedmont Ave}
  \city{Berkeley}
  \state{CA}
  \country{USA}
  \postcode{94720}
}
\author{Catherine Vera}
\authornotemark[1]
\email{cateyvera@berkeley.edu}
\affiliation{%
  \institution{University of California, Berkeley}
  \streetaddress{2224 Piedmont Ave}
  \city{Berkeley}
  \state{CA}
  \country{USA}
  \postcode{94720}
}
\author{CJ Manna}
\email{charlesmanna@berkeley.edu}
\affiliation{%
  \institution{University of California, Berkeley}
  \streetaddress{2224 Piedmont Ave}
  \city{Berkeley}
  \state{CA}
  \country{USA}
  \postcode{94720}
}


\begin{abstract}
In this study, we introduce a novel two-pronged scoring system to measure individual and systemic bias in immigration courts under the U.S. Executive Office of Immigration Review (EOIR). We analyze nearly 6 million immigration court proceedings and 228 case features to build on prior research showing that U.S. asylum decisions vary dramatically based on factors that are extraneous to the merits of a case. We close a critical gap in the literature of variability metrics that can span space and time. Using predictive modeling, we explain 58.54\% of the total decision variability using two metrics: partisanship and inter-judge cohort consistency. Thus, whether the EOIR grants asylum to an applicant or not depends in majority on the combined effects of the political climate and the individual variability of the presiding judge — not the individual merits of the case. Using time series analysis, we also demonstrate that partisanship increased in the early 1990s but plateaued following the turn of the century. These conclusions are striking to the extent that they diverge from the U.S. immigration system’s commitments to independence and due process. Our contributions expose systemic inequities in the U.S. asylum decision-making process, and we recommend improved and standardized variability metrics to better diagnose and monitor these issues.
\end{abstract}



\maketitle

\section{Introduction}
Tens of thousands of people apply for asylum in the U.S. each year, navigating an intricate and variable legal system \cite{refugeeRoulette}. This process often takes years as applications are evaluated by U.S. Citizenship and Immigration Services (USCIS), sometimes the Executive Office of Immigration Review (EOIR), and in rare cases, the Board of Immigration Appeals (BIA) \cite{typesOfAsylum}. Immigration courts are less strictly bound to precedent than domestic courts, making them more susceptible to variations in asylum decisions based on a variety of factors that are ostensibly irrelevant to the merits of the case \cite{refugeeRoulette, chenCan2017}. Prior research has exposed striking disparities in U.S. asylum grant rates as a function of the nationality of an applicant and the sociodemographic characteristics of the presiding judge \cite{refugeeRoulette}. Subsequent quantitative research has reinforced these findings by demonstrating that judicial characteristics and other extraneous features which should theoretically have no influence on an immigration judge’s decision-making are highly predictive of asylum decisions \cite{chenCan2017}.

In this paper, we seek to expand upon previous research by analyzing decision variability in U.S. asylum adjudication on two levels: We consider variability stemming from individual bias, which refers to the extent to which a particular immigration judge’s identity and experiences inform their decision-making; and partisan bias, which refers to the impact of politics on case decisions. Our study asks: \textit{How variable, if at all, are judicial decisions in asylum adjudication? What is the relationship between asylum decisions, individual judge characteristics, and systemic factors in case outcomes?}

To answer these questions, we introduce a novel framework that quantifies the notions of cohort consistency and partisanship, which represent different levels of judicial variability. Drawing on the definition of counterfactual fairness, we posit that an asylum decision is unaffected by different forms of bias if it is consistent in similar cases where particular factors such as political climate or presiding judge are perturbed.\footnote{Perturbation is the act of changing features to create counterfactuals.} We define decision variability along these two axes as proxies for bias.

\subsection{Background} \label{Background}
An asylum seeker is someone who has been displaced from their country of origin due to a “well-founded fear of being persecuted” \cite{convention_refugees} on the basis of at least one of five characteristics: race, religion, nationality, membership in a particular social group, and political opinion \cite{immigration_nationality_act}. Asylum proceedings are conducted affirmatively through the U.S. Citizenship and Immigration Services (USCIS) or defensively through the Executive Office of Immigration Review (EOIR) by pleading their case to an immigration judge \cite{affirmative, typesOfAsylum}. Our study uses data from the EOIR on defensive asylum cases.

Immigration courts cannot themselves establish precedent. The Board of Immigration Appeals (BIA) acts as the appellate court for the immigration court system and is the highest administrative authority on interpreting and applying immigration laws. The BIA’s decisions are binding on all immigration courts and DHS officers “unless modified by the Attorney General or a federal court.” \cite{bia_board} Thus, judges at the immigration court level are expected to regularly review the BIA’s decisions to ensure their decisions are consistent with the BIA’s precedent, and thereby each other.

\subsection{Contributions}
We make three novel contributions to this area:
\begin{enumerate}
\item We introduce a scoring mechanism  to represent inter-judge cohort consistency and political partisanship. This mechanism repurposes the notion of counterfactual fairness used in machine learning to operationalize the measurement of judicial decision variability in a legal context.
\item We leverage predictive modeling in asylum data to explain asylum decision variability based on cohort consistency and partisanship. This represents a departure from traditional methods of studying the qualities of asylum seekers to examining the immigration court system itself.
\item We find that partisanship explains a majority (58.54\%) of the variability in asylum decisions. This emphasizes the unique vulnerability of the immigration court system to changes in political climate.
\end{enumerate}

\section{Literature Review}
Foundational studies on asylum decision variability have exposed significant disparities in grant rates based on the applicant’s legal representation and sociological characteristics of the presiding judge, including their gender and type and length of work experience \cite{refugeeRoulette}. Among other trends, Ramji-Nogales, Schoenholtz, and Schrag’s foundational 2007 book \textit{Refugee Roulette} stated their findings that women judges granted asylum to 44 percent more cases than their male counterparts; and applicants were statistically more likely to receive asylum if their assigned judge previously worked in a private immigration law firm, worked at a nonprofit organization, or taught as a full-time law instructor \cite{refugeeRoulette}. In 2016, the U.S. government’s internal review by the Government Accountability Office (GAO) confirmed similar inter-judge variability even when controlling for the particular immigration court and type of case. In particular, the GAO replicated the aforementioned findings that the gender of a judge and the length of time a particular judge had been in office were significant predictors of asylum outcomes — judges were statistically more likely to grant asylum if they were women or had been an immigration judge for more than 3.5 years \cite{us_government_accountability_office_asylum_2016}.

Recent studies have sought to further characterize the occurrence of inter-judge variability using a variety of techniques including predicting the probability of an asylum decision and comparing actual judicial decisions to predictions made by a model trained on data that excludes the judge \cite{kamalaharan_predicting, danziger_extraneous_2011}. Notably, some studies attribute the variability to the gambler’s fallacy \cite{chen_decision-making}.\footnote{The belief that if something has previously happened more frequently than average it will happen less frequently than average in the future to balance out, or vice versa.}

A key goal of this existing literature has been predicting judicial decisions and identifying important features used to make those predictions. Past studies have achieved a prediction accuracy for grant decisions at the immigration court level between 80\% and 87\%, depending on the technique and specific data subset employed. In doing so, researchers have also investigated the impact of different extraneous factors, such as the history of a judge’s decisions, the asylum seeker’s country of origin, football scores from the previous day, and weather \cite{chenCan2017, kamalaharan_predicting}. Similar analyses have been conducted beyond U.S. immigration courts, with comparable results in data from the Board of Immigration Appeals and in international immigration courts \cite{chen_machine_predict, danziger_extraneous_2011}. Other researchers have focused on pre-trial predictability to enable attorneys to distribute resources most effectively for a case, and those models had slightly lower accuracies at 71\% to 81\% \cite{dunn_early_2017}.

An increased application of machine learning to social contexts has prompted greater scrutiny of prediction fairness and the development of a variety of metrics to measure it. Common strategies include equalized odds\footnote{The probability of a data point receiving each prediction is equal across the different values of a sensitive attribute (e.g. gender).} and statistical parity \cite{mehrabi_survey_2022}.\footnote{The likelihood of a data point receiving a positive prediction is equal for those who are and are not part of a protected group (e.g. nonbinary).} Another conception of fairness is counterfactual fairness, which analyzes fairness between members of a group by determining whether an individual’s case outcome would remain the same had they been a member of a different group \cite{kusner_counterfactual_2018}. Bias has been quantified in more granular ways as well: in the field of algorithmic auditing, partisanship has been formalized as a probability of party affiliation of search snippets \cite{hu_auditing_2019}.

Scholars have identified a variety of factors that influence partisanship in asylum decisions: the immigration judge’s prior professional experience \cite{refugeeRoulette}, public sentiment toward immigration \cite{johnson_isaiah_under_2019}, the political party in power \cite{yarnold_barbara_m_federal_1990, brace_measuring_2000}, U.S. foreign policy, U.S. relations with the country of asylum-seeker’s origin, \cite{salehyan_international_2008}. Empirical studies on politicization of immigration judges regarding orders of removal reveal a statistically significant relationship between the administration in power and an immigration judge’s decision \cite{kim_empirical}.

Existing literature quantifying judge fairness has distinguished between two broad categories: procedural, referring to fairness of the procedures used to reach decisions, and distributive, referring to fairness of outcomes \cite{tyler_role_1984}. Some scholars conflate distributive fairness with justice, which makes the former difficult to quantify. Much of the focus in the literature is on procedural fairness, which focuses on the methods judges use to implement the law, such as safeguards for asylum applicants against strict time constraints on their cases \cite{cropanzano_measuring_2015, ippolito_recast_2011}.

\subsection{Gaps in the Literature}
To build upon prior literature, we seek to formalize a method to measure judicial decision variability within immigration courts. While prior research has quantified fairness in federal courts, it has centered on proceedings under the judicial branch of the U.S. government. We seek to expand this to immigration courts, which abide by a different set of rules as part of the executive branch. Though existing research has investigated decision variability in asylum through odds ratios and other techniques \cite{us_government_accountability_office_asylum_2016}, this notion has not been expressed in a manner that can be compared across spatial and temporal boundaries. We present a novel two-pronged scoring system to address this need.

A secondary objective of our work is to generate new insights from our data via predictive modeling. Previous studies analyzing asylum data with predictive modeling seek to predict proceeding outcomes with pre-decision data. This approach reinforces the traditional role of predictive modeling as a future-facing, target variable-oriented method. We seek to dismantle this in two ways: First, we include judicial decision as one of many covariates used to predict other variables of interest. Second, we adopt existing machine learning metrics to evaluate how well our scoring system captures decision variability.

\section{Methodology}

\subsection{Dataset}
To construct our dataset, we gathered information from two sources: the U.S. Executive Office of Immigration Review’s (EOIR) public dataset,\footnote{This data is available at https://www.justice.gov/eoir/foia-library-0 due to repeated Freedom of Information Act requests. We downloaded the entire dataset in January 2022.} which contains information on case proceedings, asylum seeker representation, and juvenile applicants; and the Transactional Records Access Clearinghouse (TRAC) website,\footnote{We used Python’s Selenium and Requests libraries to build a web scraping tool to retrieve each immigration judge’s online biography from the TRAC website. It collects links for judges who were active between 2006 and 2021 and scraped the associated HTML and text biographies for the valid links.} which features detailed biographies on immigration judges. We limited our dataset and analysis to cases charged before January 2022 and after the implementation of the Refugee Act of 1980. The data was cleaned and merged along proceeding ID number and further features such as a judge's monthly decision count were computed and added to the dataset.

Our dataset faces two primary limitations: First, we do not have access to data on the merits of an asylum case, which includes the grounds for receiving asylum and justification for why an asylum seeker’s “subjective fear of persecution” is “objectively reasonable” (see Section \ref{Background}). Second, our results risk reflecting survivorship bias: Due to the immense case backlog, many recently charged cases do not have a decision yet; since our research focuses on decision variability, we removed pending proceedings from the dataset. The ones that are remaining from 2020 and 2021 have been processed more quickly than average.

The final dataset contains 5,975,440 rows which correspond to proceedings and 234 columns which correspond to information about each proceeding. Of these features, 83 features contain information on case proceedings (such as charging dates and decision outcomes), 36 features contain information on the asylum seeker’s background (such as local residence and demographics), 87 features contain information on the presiding judge (such as professional background and experience), and 28 features serve as null indicator columns.

\subsection{Variability Scoring}
While many formalizations of fairness have been developed to study machine learning models \cite{mehrabi_survey_2022}, our analysis seeks to evaluate decision fairness in immigration courts. As such, we are subjected to contextual constraints. For example, different nationality groups in the dataset are expected to have different grant rates. Since the particular conditions of their country of origin may determine their eligibility for asylum, we are not interested in equality between them. As such, group fairness\footnote{Each group has a similar probability of an outcome.} would be inappropriate. On the other hand, the legal context of this data emphasizes the relevance of individual fairness.\footnote{Individuals in a group have similar outcomes.} The precedent set by the Board of Immigration Appeals as well as guidelines from the attorney general indicate that case decisions should abide by a consistent set of standards. In other words, one can expect similar outcomes for similar cases, as judges ought to apply the court-wide standards similarly and come to similar decisions. We thus draw upon counterfactual fairness to quantify decision variability: A decision is fair if it would have been consistent in a world where a demographic attribute of an individual is changed but all else is equal \cite{kusner_counterfactual_2018}. In our case, we are interested in quantifying decision variability by perturbing particular attributes of a case and measuring its effects on an immigration judge’s decision.

Traditionally, algorithmic fairness methods compare a model’s decision to the correct one provided in the data. As there is no objectively “correct” decision about whether a given asylum seeker should receive asylum, we employ counterfactual analysis to infer decision accuracy. We compare each decision to those made for similar cases, which we define as sharing the following three attributes: asylum seeker nationality, which affects their need for asylum since they are requesting to permanently leave their country; the primary immigration court where the trial is held, since cases are distributed randomly among judges within these \cite{reuters_not-quite-independent_2017}; and the time interval (in 5-year bins) that the decision was made in, due to autocorrelation observed in prior literature indicating that temporality is a significant factor \cite{chen_proceedings_2017}. As part of the executive branch, immigration courts have the flexibility to respond to global changes such as refugee crises. We chose to hold these features constant to avoid a scoring penalty for such behavior.

Thus, we created two novel scoring metrics to computationally measure both individual and systemic variability. Specifically, we sought to quantify individual judicial variability relative to their fellow judges, which we refer to as “cohort consistency”, and systemic impacts of the political zeitgeist on judicial decisions, which we will refer to as “partisanship.”

\subsubsection{Individual Variability: Cohort Consistency}
According to the U.S. Justice Department’s code of conduct for immigration judges, “An Immigration Judge who manifests bias or prejudice in a proceeding impairs the fairness of the proceeding and brings the immigration process into disrepute" \cite{professionalism_guide}. While we seek to avoid making value judgements regarding the biases of individual judges, we hope to form a generalized understanding of how individual biases may affect immigration decisions in general. As such, we are interested in measuring the cohort consistency of individual judges, as determined by the similarity of decisions made between them.

Based on the theory of counterfactual fairness, we stipulate that \textit{a given judge is consistent if, on average, different judges would have ruled on each of their cases in the same way}. We measure decision agreement between judges who rule on cases from the same cohort.\footnote{We define this as the subset of cases with the same immigration court, nationality, and timeframe.} Thus, for each case attributed to each judge, we compute the proportion of other judges who agree with the decision of the judge in question. After calculating this, we average out the agreement score for each of the judge’s decisions to determine their overall cohort consistency score.

\paragraph{Cohort Consistency Formula}
First, let us define four values: $\Delta_{fi}$ is the value of feature $fi$ for the asylum proceeding in question $\Delta$. $\delta_{fi}$ is the value of feature $fi$ for all other proceedings $\delta$. $\sigma$ is the set of all asylum proceedings. Then, we can define the set of similar proceedings $S_\delta$ used to build counterfactuals as follows: For each feature $fi$ among those held constant for similarity, $\delta_{fi}$ has the same value as $\Delta_{fi}$.

\begin{equation} \label{Similar Proceedings}
  S_\delta := \sigma\{\delta | \delta_{NAT} = \Delta_{NAT}, \delta_{IC} = \Delta_{IC}, \delta_{YR\_BN} = \Delta_{YR\_BN}\}
\end{equation}

Then, the set of all counterfactuals of the proceeding in question with the judge feature perturbed is given by the following:

\begin{equation}
    Proc_\delta := S_\delta\{\delta | \delta_{JUD} \neq \Delta_{JUD}\}
\end{equation}

This helps us define the proportion of judges who agree with the decision of the judge on the case in question.

\begin{equation}
    \Omega_\delta := Pr(Proc_\delta\{\delta | \delta_{DEC} = \delta_{DEC}\})
\end{equation}

Finally, we have the overall fairness score of the judge, based on all their decisions, as follows:

\begin{equation}
    \Phi_{JUD} := \frac{\sum_{\delta \in Proc} \Omega_\delta}{|Proc|}
\end{equation}

\paragraph{Interpretation} A score close to 1 indicates that a judge frequently agrees with other judges on similar cases. A score close to 0 indicates that they frequently disagree. For example, a cohort consistency score of 0.85 signifies that, on average, a judge agrees with 85\% of other judges who ruled on similar cases.

\subsubsection{Systemic Variability: Partisanship} \label{Systemic Variability: Partisanship}
We are defining “partisanship” as the extent to which an asylum decision is impacted by the political climate in which it was made. Formalizing this in terms of counterfactuals, we state that \textit{a decision is partisan if changing the political climate in which it was made results in the opposite decision}. Note that this is politically neutral; there is no definitional association between political party and score. While political climate is a broad concept, we choose to represent it using two factors: the political administration currently in power in the presidency, and the political leaning of the state as evidenced by the party that received the majority vote in the last presidential election.

\paragraph{Partisanship Formula}
We define the set of similar proceedings $S_\delta$ in the same manner as Equation \ref{Similar Proceedings}. We also define the set of all proceedings with political climates that are different from that of the proceeding in question. This definition involves a different political party in power or a different political leaning of the state. Thus, the set of all counterfactuals of the proceeding in question with the political climate features perturbed is given by the following:

\begin{equation}
    Proc_\delta := S_\delta\{\delta | \delta_{NAT\_POL} \neq \Delta_{NAT\_POL} || \delta_{ST\_POL} \neq \Delta_{ST\_POL}\}
\end{equation}

This helps us define the proportion of opposing decisions under a different political climate, representing the level of partisan influence on the current decision.

\begin{equation}
    \Gamma_\delta := Pr(Proc_\delta\{\delta | \delta_{DEC} = \delta_{DEC}\})
\end{equation}

\paragraph{Interpretation} A score close to 0 indicates that a decision is frequently the same as those made for similar cases in different political climates. A score close to 1 indicates that it is frequently different. For example, a partisanship score of 0.85 signifies that 85\% of similar cases made in different political climates had the opposite decision.

\subsubsection{Limitations}
Counterfactual fairness assumes that we have access to all the same information as the decision-maker (in the case of machine learning, the model), but in our analysis we do not since we lack the arguments made by each side; however, our application of the method circumvents this by considering only four variables in total. When using counterfactual proxies, moreover, there is the risk that scores do not accurately reflect what “partisanship” and “cohort consistency” actually mean. Due to its reliance on similarity, the scoring mechanism would result in high cohort consistency scores and low partisanship scores in places like El Paso, Texas where asylum is rarely granted \cite{trac_tool}. These scores do not capture whether the decision \textit{ought} to have been made in this way and thereby discounts these notions of fairness and justice. Furthermore, the definition of political climate used when calculating partisanship considers only the political party of the sitting president and the political leaning of the state. We found these sufficient because they provide a view into current political trends at both the state and federal levels, but there are many other factors that could have been considered in their place.

\subsection{Variable Correlations}\label{Variable Correlations}
This study uses predictive modeling as a way to understand the significance of different factors in influencing various outcomes of interest. Our approach assumes that the relationship between variables in the dataset can be learned from the structure of an accurate model, a tenet of the data-modeling culture of statistics \cite{breiman_statistical_2001}.\footnote{The data-modeling culture assumes a stochastic relationship between variables in the data.} Since this culture has been criticized for limiting the scope of questions researchers explore and leading to fallacious conclusions, we instead blend techniques to focus on tree-based machine learning methods such as random forest associated with the algorithm-modeling culture, its alternative \cite{breiman_statistical_2001}.\footnote{Ignores the relationship between variables; more focused on accurate prediction.}

We examined covariates that affected a selection of variables we chose based on existing literature and our research question. Due to our interest in political climate, we predicted the political party of the sitting president at the time of the decision. Since presidents set the national agenda and influence immigration policy and regulations, as described above, we believe that this is an important variable to consider. Different parties also have different views regarding treatment of asylum seekers, so this analysis would investigate that as well. Additionally, since our study focuses on asylum decision variability, we predicted the partisanship and cohort scores produced using the methods described earlier to uncover which variables correlate with them.

\subsubsection{Model Construction}
Our analysis was conducted in Python, using Pandas \cite{reback2020pandas, mckinney-proc-scipy-2010} to store and manipulate data and Scikit-Learn \cite{scikit-learn} to build models.

Before building models, we preprocessed our data: First, we applied frequency encoding\footnote{Replaces each occurrence of a string with its overall frequency, as a decimal between 0 and 1, within the dataset.} to categorical variables. Next, we removed rows where our target variable was null. Finally, due to high correlation between variables in the dataset, we selected features for models by removing one of each pair of features for which the correlation is greater than 0.95. This allows us to avoid representing information redundantly and provides more interpretable results about which variables are important in prediction.

We built random forest classifiers and regressors as appropriate for each variable. To extract insights from our models, we analyzed feature importances. In acknowledgement of the criticism of the data modeling paradigm mentioned previously, we used bagging to compute means and standard deviations of feature importances. We then computed Spearman’s correlation coefficient for nonlinear relationships to determine effect size and statistical significance, which was examined after applying a Bonferroni correction.

\subsubsection{Limitations}
The main limitation of our methodology was computational resources. Attempts to increase accuracy by employing the common practice of data standardization\footnote{Subtracts the means and scales each feature to a unit variance.} drastically increased the size of the data from 3.5 GB to over 10 GB, leading us to forgo the procedure. Additionally, when bootstrapping the random forests 1,000 times, we set certain model hyperparameters to lower values to maximize performance. Each random forest was trained on 5,000 data points (about 0.1\% of the dataset) with each tree within that fit to a maximum sample size of 1,000 data points.

\subsection{Decision Prediction}
This analysis was conducted using Python, Pandas, and Scikit-Learn. We recognize that the data we have on asylum seekers has been extracted from people in vulnerable circumstances. Thus, this study uses the data to critically examine the asylum system rather than the people that have been subjected to it; thus, we focus on how well our computational measures of individual and systemic outcome variability capture judicial decisions.

In order to characterize the relationship between our variability scores and proceeding decisions, we constructed models that predict decisions using partisanship, cohort consistency, and disaggregated consistency. We used model performance metrics to evaluate how much variance in the data is captured by these scores; this represents the amount that a decision was influenced by variability in asylum processes.

\subsubsection{Model Construction}
We performed this analysis on each valid score combination in our data: partisanship and cohort consistency, partisanship and disaggregated consistency, solely partisanship, solely cohort consistency, and solely disaggregated consistency. Since variables for political climate and judge identity sometimes contained null values, partisanship and consistency scores were not available for every point in our dataset. Therefore, each model was trained on 80\% of the data that existed with the given scores and evaluated on the remaining 20\%.

We trained two models on each subset of data. One was standard logistic regression. Observing that a linear boundary may be appropriate for our dataset, we also implemented linear support vector classifiers. Our models were evaluated on two main metrics: The test accuracy helps to legitimize our scoring system by determining whether the model is appropriate, and the R$^2$ captures how much variability in decisions can be explained by the set of scores used for prediction.

\subsubsection{Limitations}
A significant caveat exists in the interpretatin of our results: Partisanship, cohort consistency, and disaggregated consistency are all calculated using judicial decision. This results in some leakage\footnote{When information that includes the target variable is inadvertently used to predict it.} since we are using covariates that depend on the target variable. However, the scores incorporate information about political climate and judges that are not encompassed by the decision itself. Additionally, since the scores are calculated based on agreement with the decision and not its actual value (1 for grant, 0 for denial), we avoid using it directly.

\subsection{Time Series Analysis}
We believe useful inferences about the influences of specific political statutes and events can be extracted using time series analysis. Our analysis was conducted in Python, using Pandas to store and manipulate data, Facebook’s Prophet package \cite{taylor_forecasting_2017} to model our data, and Matplotlib \cite{hunter_matplotlib_2007} to visualize trends.

\subsubsection{Model Fitting and Trend Analysis}
Using Prophet, we fit a decomposable additive functional model to our data, aggregated by week, with two main model components: trend and seasonality within a year. Our trend model is a piecewise constant function with a number of changepoints where the rate of growth can be changed. Seasonalities are estimated by Prophet using partial Fourier sums.

The rate of growth is fit and modified at changepoints using Prophet’s automatic changepoint detection. It first specifies a large number of potential changepoints in the data where the rate of growth is allowed to change, and then puts a sparse Laplace prior distribution on the magnitude of the rate changes. This has the effect of L1 regularization, such that Prophet will use as few of the changepoints as possible to fit the model’s curve to the data.

To find our model parameters, we conducted a grid search over two parameters: the scale of the changepoint prior and the scale of the seasonality prior. To get error values, we performed time-series cross-validation by setting a series of one-year hold-out segments of data, training on the data from before the initial cut-off point, and then validating the predicted values against the historical data. Our final model parameters with the lowest root mean squared error is 0.1 for the changepoint prior scale and 0.01 for the seasonality prior scale.

Once the model had been fit, we visualized the decomposition of the model in order to analyze the change in the trend across our data, as shown in Section \ref{time-series}.

\subsubsection{Limitations}
Since our case data is not evenly distributed over our period of analysis, aggregating the scores over smaller time granularities and filtering to specific nationalities left us with few cases to calculate the average over, resulting in certain cases being more heavily weighted in the aggregated data points. This is particularly relevant before 1990, where the volume of cases is much lower.

Additionally, the changepoints in our trend are initially randomly distributed in time using Prophet’s changepoint detection. Since there is no significance to the exact dates chosen for potential changepoints, changes in the rate of growth observed in the trend occur in the locality of the date plotted, but not necessarily exactly as plotted.

\section{Results}
\subsection{Dataset}
We began by examining our dataset using descriptive statistics and visualizations made with the Seaborn package \cite{waskom_seaborn_2021} to identify notable trends. Since exploratory analysis cannot be conflated with statistically significant results, we augmented this by determining variable correlations using predictive modeling as described in Section \ref{Variable Correlations}.

\begin{figure}
\caption{Volume of immigration court proceedings between 1980 and 2021.}
\includegraphics[width=\linewidth]{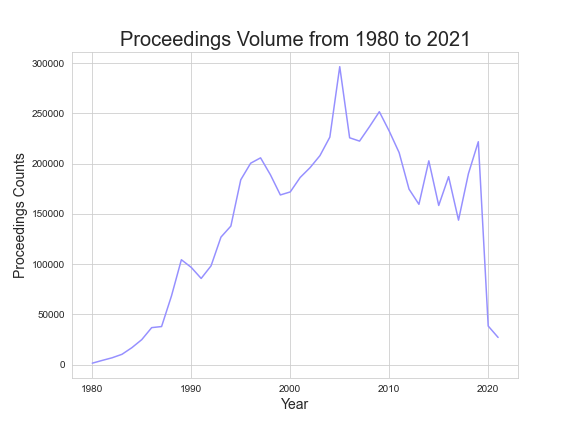}
\centering
\label{fig:proceedings_year}
\end{figure}

\begin{figure}
\caption{Annual asylum grant rate between 1980 and 2021.}
\includegraphics[width=\linewidth]{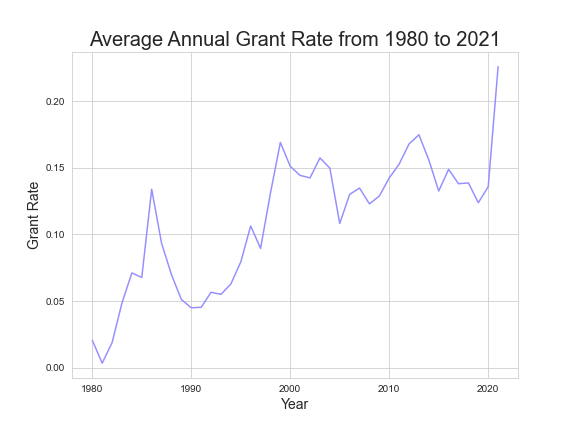}
\centering
\label{fig:grant_rate_year}
\end{figure}

Figure \ref{fig:proceedings_year} depicts the number of decided proceedings in our dataset each year between 1980 and 2021. Notably, the annual case volume peaked at 296,485 in 2005. We observed an annual average of 142,270 cases since 1980. Likewise, Figure \ref{fig:grant_rate_year} shows the percentage of granted applications for each year in our dataset over the same timeframe. The most significant valley in grant rate occurred around 1990 at just under 5\%. The average grant rate over the full dataset is 12.875\%. Though grant rate appears to have peaked in 2021, this is associated with a significant drop in case volume, suggesting that a number of cases from that year are still undecided being processed in the backlog at the time of writing.

\begin{figure}
\caption{Proportion of proceedings where asylum seekers received representation between 1980 and 2021.}
\includegraphics[width=\linewidth]{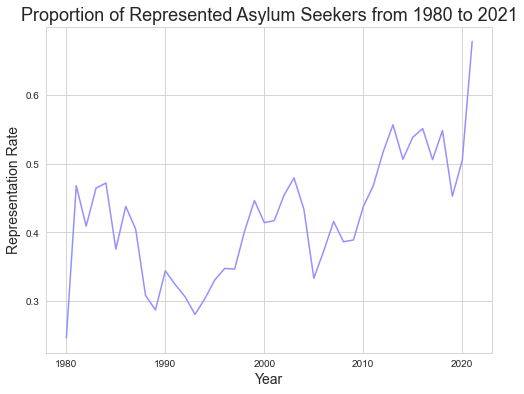}
\centering
\label{fig:representation_year}
\end{figure}

Figure \ref{fig:representation_year} shows the percent of proceedings where an asylum seeker had attorney representation each year between 1980 and 2021. Notably, the annual representation proportion peaked at 55.66\% in 2013 and again at 67.78\% in 2021. 43.07\% of proceedings since 1980 have had attorney representation for the asylum seeker.

\begin{figure}
\caption{Breakdown of custody status for asylum seekers with and without representation.}
\includegraphics[width=\linewidth]{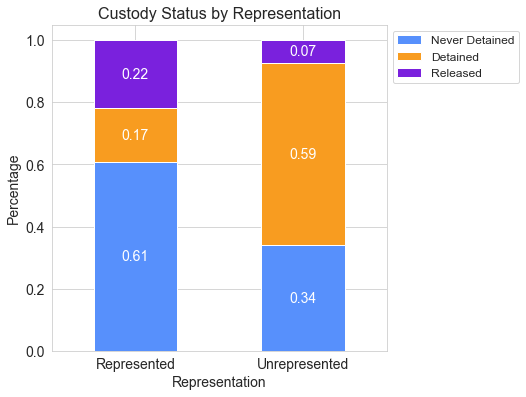}
\centering
\label{fig:representation_custody}
\end{figure}

Notably, 26.308\% of represented asylum seekers received asylum, while only 2.712\% of unrepresented asylum seekers did. Figure \ref{fig:representation_custody} illustrates similarly large disparities in custody status: 58.695\% of asylum seekers without representation were detained, compared to only 17.442\% of those with representation.

\begin{figure}
\caption{Proceeding duration for represented and unrepresented asylum seekers.}
\includegraphics[width=\linewidth]{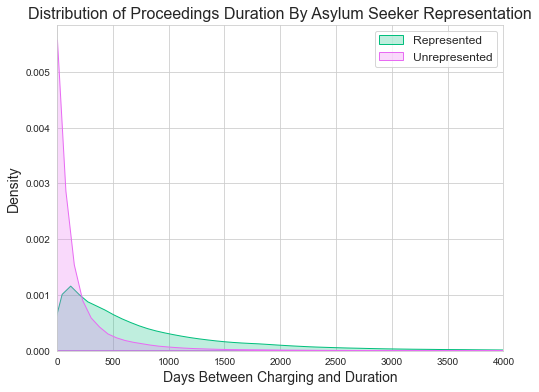}
\centering
\label{fig:representation_decision_time}
\end{figure}

Figure \ref{fig:representation_decision_time} visualizes the change in time, in days, between the charge and the decision of each proceeding. We observed that the majority of durations for unrepresented asylum seekers tends to be concentrated below 500 days, while that of represented asylum seekers is below 1000 days. The averages reflect this disparity: proceedings take an average of 825 days for unrepresented asylum seekers and 904 days for represented asylum seekers.

\subsection{Cohort Consistency Scores}
Next, we analyzed associations between cohort consistency and other variables. The average cohort consistency score of the dataset is 0.790468.

\begin{figure}
\caption{Distribution of grant rate compared to cohort consistency.}
\includegraphics[width=\linewidth]{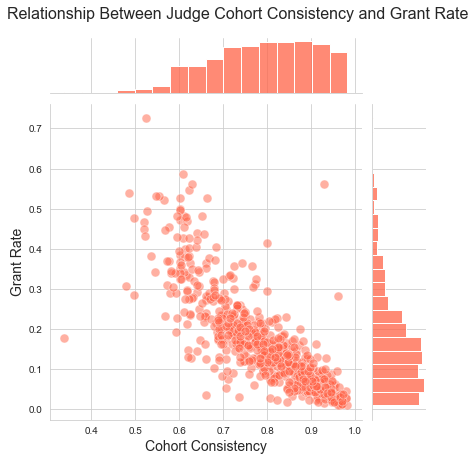}
\centering
\label{fig:consistency_grant_rate}
\end{figure}

Figure \ref{fig:consistency_grant_rate} visualizes the relationship between the judge cohort consistency score and asylum grant rate. We identified a strong negative correlation between the two variables: the standardized covariance is -0.81787. Judges with high cohort consistency scores tend to display a low grant rate.

We next stratified cohort consistency scores by various characteristics of judges. Prior literature suggested a relationship between gender and grant rate, so we examined its impact on consistency. Additionally, due to our general interest in partisanship, we analyzed the political party of the administration that appointed the judge. 

\begin{figure}
\caption{Distributions of cohort consistency stratified by judge gender.}
\includegraphics[width=\linewidth]{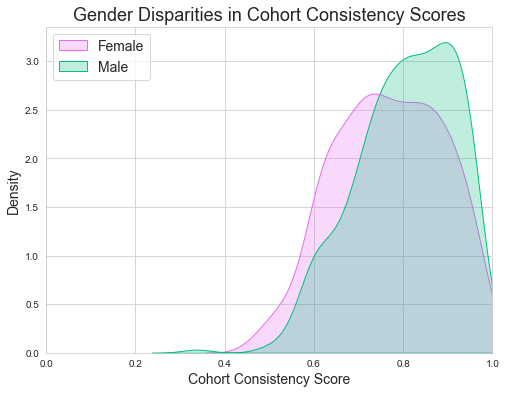}
\centering
\label{fig:consistency_gender}
\end{figure}

\begin{figure}
\caption{Distributions of cohort consistency stratified by appointing political party.}
\includegraphics[width=\linewidth]{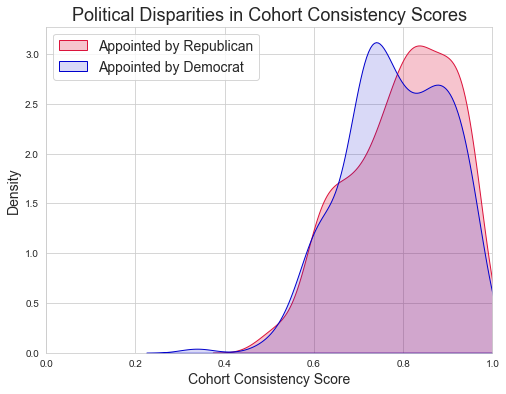}
\centering
\label{fig:consistency_party}
\end{figure}

Figure \ref{fig:consistency_gender} displays the relationship between cohort consistency scores and the gender of the judge. We observe that the average consistency score of female judges is 0.76695, compared to 0.804427 for male judges. Figure \ref{fig:consistency_party} performs the same analysis for the political party of the administration admitting the judge. There is a negligible difference between the cohort consistencies of judges appointed by Democrats and by Republicans, with a mean value of 0.78217 and 0.796202 respectively.

The model to predict our measure of cohort consistency had a RMSE of 0.11745. The R$^2$ of the model was 0.04838 and the R was 0.21995. Given that this model’s statistics are quite poor, the following associations should be viewed with a degree of skepticism.

\begin{table}
\caption{The dataset variables associated with higher cohort consistency scores.}
\label{tab:variables high consistency}
\begin{tabular}{@{}lllll@{}}
\toprule
\textbf{Feature}               & \textbf{Mean} & \textbf{StD} & \textbf{Coef} & \textbf{P} \\ \midrule
Male                    & 0.022                & 0.008               & 0.204               & 0.005019            \\
Member of TX Bar                 & 0.024                & 0.010               & 0.193               & 0.007902            \\
Member of GA Bar                 & 0.008              & 0.006               & 0.176               & 0.015907            \\
Member of LA Bar                 & 0.012                & 0.008               & 0.168               & 0.021331            \\
Appointed by Att. Gen. Meese & 0.002                & 0.002               & 0.160               & 0.028585            \\ \bottomrule
\end{tabular}
\end{table}

Table \ref{tab:variables high consistency} describes the bagged values and significance of variables with higher cohort consistency Spearman coefficients. After applying a Bonferroni correction for multiple comparisons, none of the features were significant.

\begin{table}
\caption{Dataset variables with the strongest negative association with inter-judge cohort consistency.}
\label{tab:variables low consistency}
\begin{tabular}{@{}lllll@{}}
\toprule
\textbf{Feature}     & \textbf{Mean} & \textbf{StD} & \textbf{Coef} & \textbf{P} \\ \midrule
Member of NY Bar       & 0.056                & 0.02               & -0.272              & 0.0**          \\
Female        & 0.025                & 0.01                & -0.242              & 0.001*           \\
Member of CA Bar       & 0.042                & 0.013               & -0.222              & 0.002            \\
Year Received Law Degree        & 0.198                & 0.022               & -0.125              & 0.086            \\
Teaching Experience & 0.018                 & 0.008               & -0.112              & 0.101            \\ \bottomrule
\end{tabular}
\end{table}

Table \ref{tab:variables low consistency} displays the same for the variables associated with lower cohort consistency scores. Two variables were found to be significant with the correction: JUDGE\_NY\_BAR under $\alpha = 0.05$ and JUDGE\_FEMALE under $\alpha = 0.1$.

\subsection{Partisanship}
We then analyzed partisanship scores on the dataset. The average partisanship score of the dataset is 0.179371.

\begin{figure}
\caption{Annually aggregated partisanship scores and grant rates.}
\includegraphics[width=\linewidth]{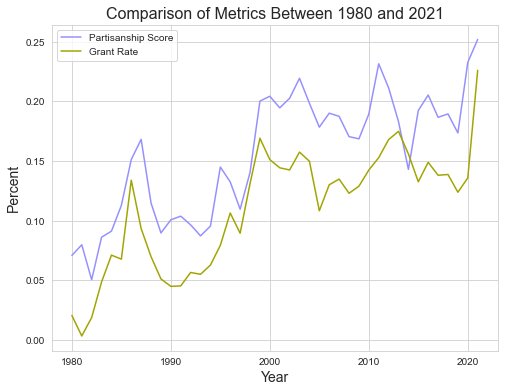}
\centering
\label{fig:scores_year}
\end{figure}

Figure \ref{fig:scores_year} displays a time series chart of the average partisanship score and asylum grant rates by year between 1980 and 2021. We observe similar behavior of the two metrics despite their distinct calculation, suggesting a high correlation. 

\begin{table}
\caption{Dataset variables with the strongest positive association with partisanship.}
\label{tab:variables high partisanship}
\begin{tabular}{@{}lllll@{}}
\toprule
\textbf{Feature}           & \textbf{Mean} & \textbf{StD} & \textbf{Coef} & \textbf{P} \\ \midrule
Decision                   & 0.630             & 0.014          & 0.548           & 0                   \\
Charging to decision (days)    & 0.008          & 0.001         & 0.443           & 0                   \\
First choice deport. country (null)   & 0.0002          & 7.09E-05               & 0.439           & 0                   \\
Alien Primary Attorney(s)       & 9.29E-04                & 3.98E-04               & 0.435           & 0                   \\
Alien Non-Primary Attorney(s) & 0.0011          & 0.0003        & 0.423           & 0                   \\
Latest Hearing is Individual   & 0.0006           & 0.0002        & 0.407           & 0                   \\
First to Last Hearing (days)   & 6.35E-03                & 0.0012         & 0.372           & 0                   \\
Attorney Number (legacy)          & 0.0017          & 0.00038        & 0.326           & 0                   \\
Base City Code                 & 0.024           & 0.0039         & 0.313           & 0            \\
Alien State Voted Democratic     & 0.001          & 3.86E-04               & 0.308           & 0            \\ \bottomrule
\end{tabular}
\end{table} 

The model to predict our measure of partisanship had a RMSE of 0.10233, an R$^2$ of 0.82608, and an R of 0.90889. Table \ref{tab:variables high partisanship} denotes the top 10 out of 107 total variables with a statistically significant positive Spearman correlation with our partisanship metric. Notably, high partisanship is associated with the last hearing being an individual hearing, a longer time between the first and last hearings, and being granted asylum.

Table \ref{tab:variables low partisanship} denotes the top 10 out of 71 total variables with a statistically significant negative Spearman correlation with our partisanship metric. Notably, our model found that cases that were less partisan were associated with an asylum seeker being detained at the time of the decision, originating from a Latin American country, and being tried in a state that voted Republican in the previous presidential election.

\begin{table}
\caption{Dataset variables with the strongest negative association with partisanship.}
\label{tab:variables low partisanship}
\begin{tabular}{@{}lllll@{}}
\toprule
\textbf{Feature}                               & \textbf{Mean} & \textbf{StD} & \textbf{Coef} & \textbf{P} \\ \midrule
First choice deport. country                            & 0.0064          & 0.0021         & -0.479          & 0                   \\
Judge decisions by month                        & 0.0163           & 0.0028         & -0.451          & 0                   \\
Currently detained                                     & 0.0007          & 2.64E-04               & -0.369          & 0                   \\
Alien nationality Central America  & 3.50E-02                & 1.04E-02               & -0.329          & 0                   \\
Alien nationality                                    & 0.0145           & 0.0037         & -0.305          & 0                   \\
Language                                       & 0.0074          & 0.0030         & -0.300          & 0                   \\
Judge Code                                    & 6.52E-03                & 0.0009        & -0.295          & 0                   \\
Base City State Voted Republican                          & 0.0012          & 0.0004        & -0.293          & 0                   \\
Alien Country Percent Christians                        & 0.0373           & 0.0091         & -0.279          & 0            \\
Decision by EOIR-7 deportation                            & 9.08E-05                & 6.37E-05               & -0.270           & 0            \\ \bottomrule
\end{tabular}
\end{table}

\subsubsection{Time Series Analysis} \label{time-series}

In order to better understand decision partisanship along a temporal axis, we conducted time series analysis on the scores between 1980 and 2021.

\begin{figure}
\caption{Time series of partisanship scores and their rolling mean between 1980 and 2021.}
\includegraphics[width=\linewidth]{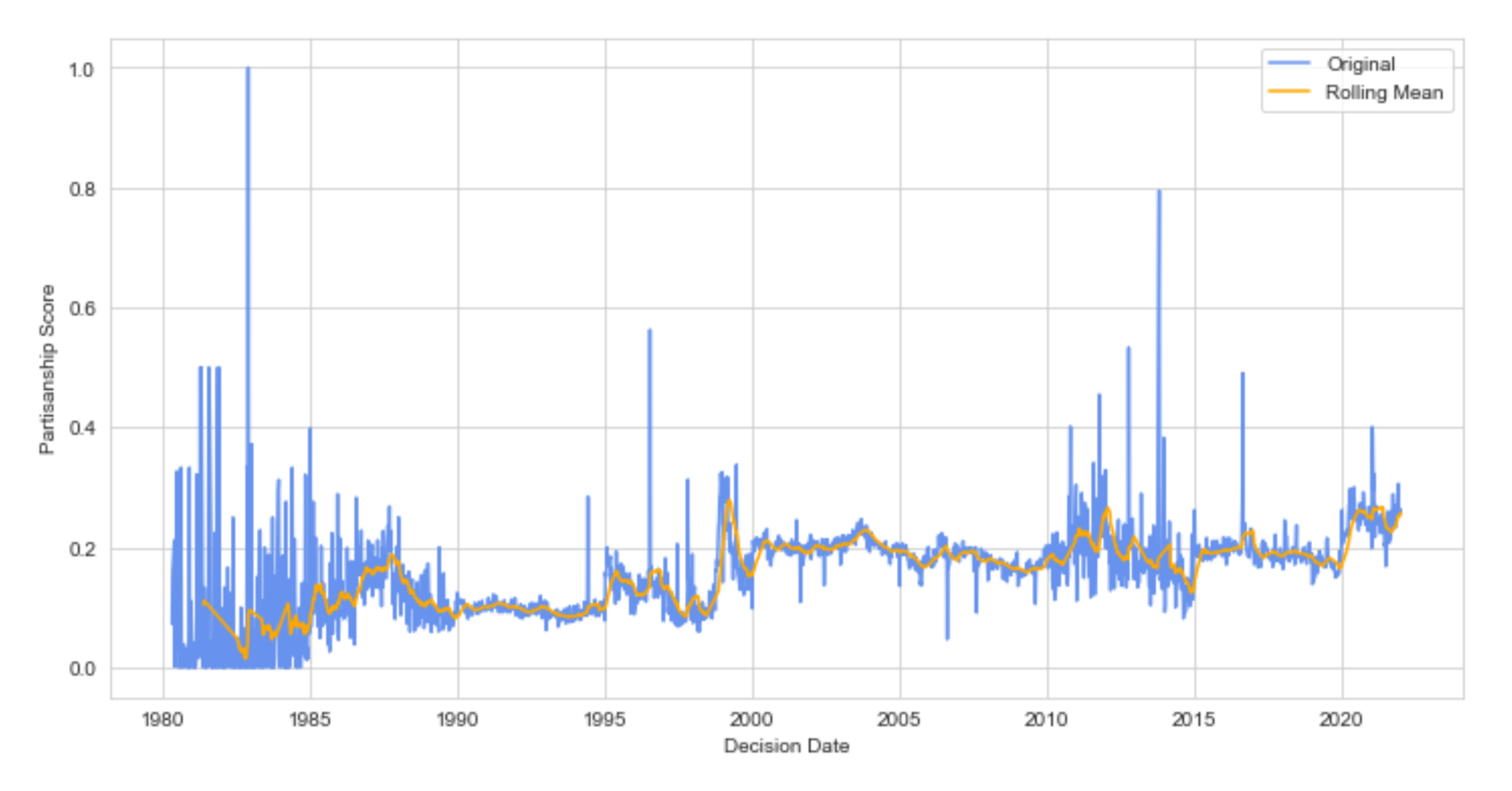}
\centering
\label{fig:time_series}
\end{figure}

Figure \ref{fig:time_series} depicts the change in partisanship scores over time aggregated by week overlaid with a plot of the rolling mean of the previous 12 data points. By visual inspection, we note greater volatility in partisanship scores in 1980-1990 and 2010-2015. We analyze a Prophet model for further insight on trends and seasonality.

\begin{figure}
\caption{Final fitted model curve with training points.}
\includegraphics[width=\linewidth]{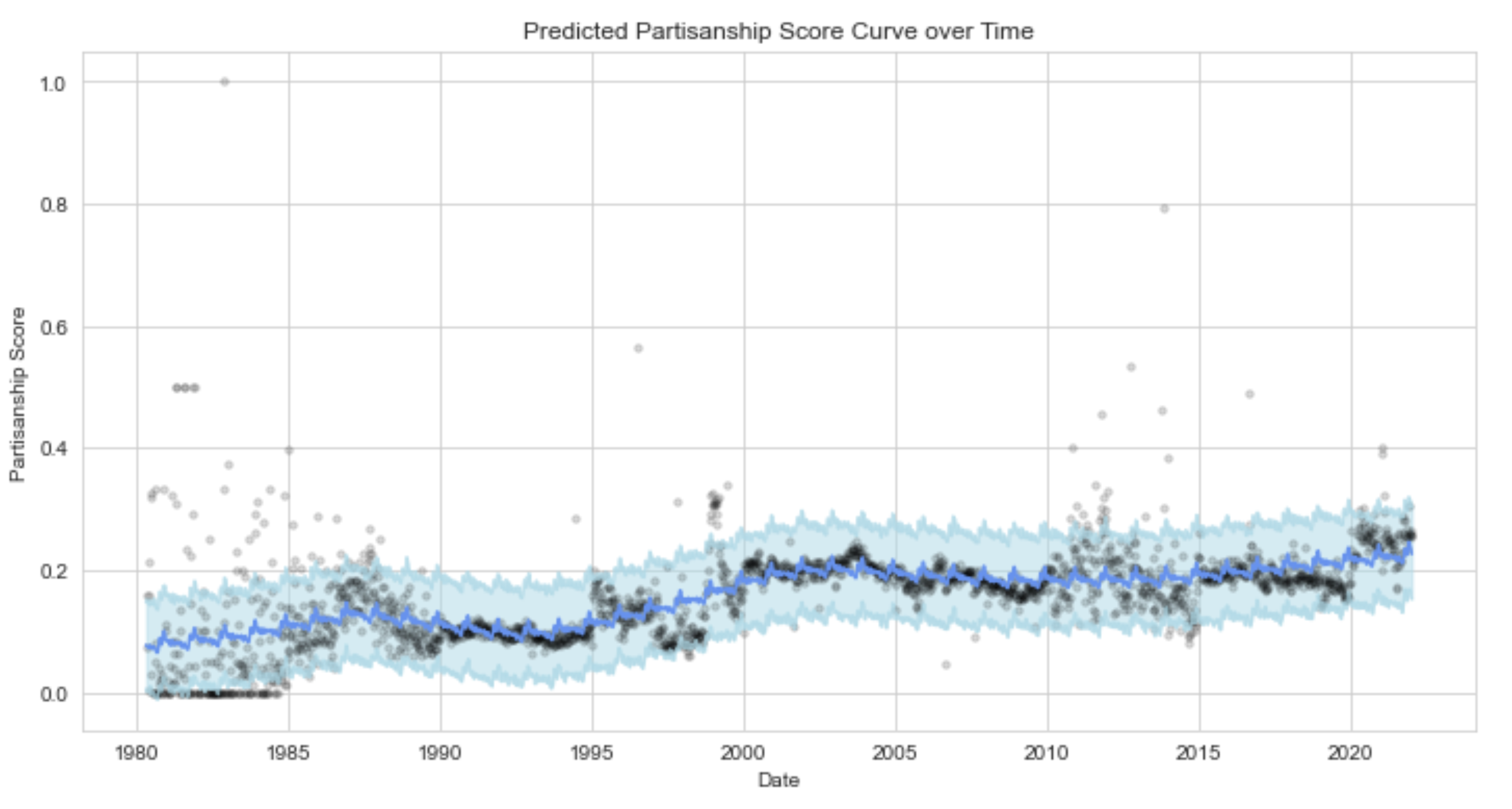}
\centering
\label{fig:fitted_time_series}
\end{figure}

Figure \ref{fig:fitted_time_series} shows the predictions of our fitted Prophet model and a corresponding confidence interval of width approximately 0.17. With cross-validation, the model’s final root mean squared error was 0.053753 and final mean absolute error was 0.036483. Its R$^2$ value was 0.384826. We next decomposed the model to better understand the overarching trend and score seasonality.

\begin{figure}
\caption{Piecewise linear trend component of Prophet model.}
\includegraphics[width=\linewidth]{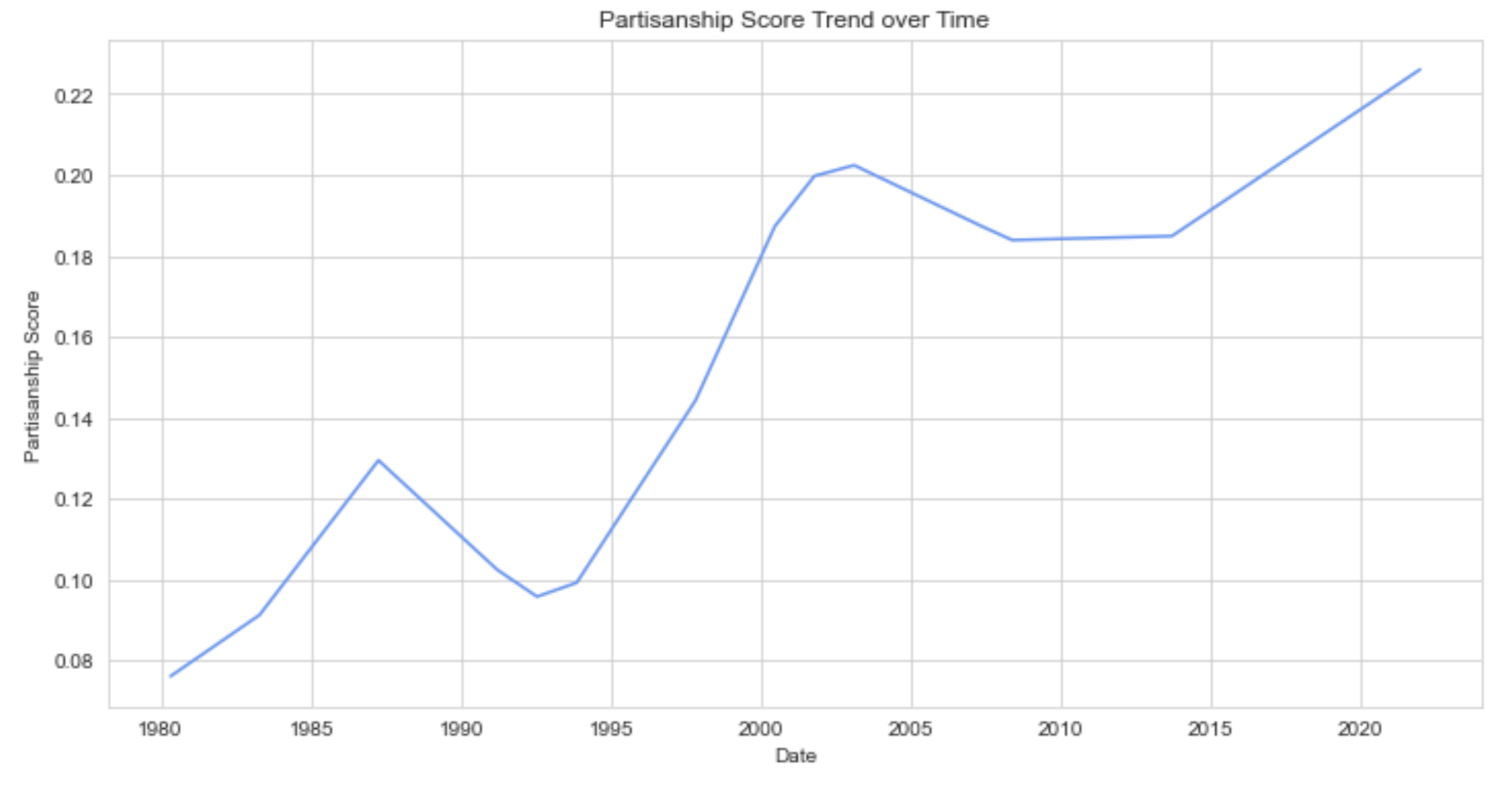}
\centering
\label{fig:piecewise_time_series}
\end{figure}

Figure \ref{fig:piecewise_time_series} displays the model’s trend component. Notably, there is a period of marked increase in partisanship score starting around 1992 and plateauing approximately after 2003. However, partisanship score rises again beginning in 2014 and continues to grow past its historical peak until the end of our data in 2022.

\subsection{Decision Prediction}
Next, we analyzed the ability of our constructed variability scores to predict asylum decisions. Our research seeks to quantify the effect of systemic and individual factors, represented by partisanship scores and cohort consistency scores respectively. In order to do so, we construct two classification models each on five different feature sets. We found that partisanship is more highly predictive of decision than cohort consistency and contributes significantly to prediction accuracy.

Before this analysis, we considered a baseline classifier that always predicts that the asylum claim will be denied. Due to class imbalance, where only 12.8751\% of cases have been granted since 1980, the accuracy of this unintelligent model is still quite high at 87.1259\%. However, the recall is 0\%. The R$^2$ value is -14.7777\%.

\begin{figure}
\caption{VDecision distribution across partisanship and cohort consistency scores.}
\includegraphics[width=\linewidth]{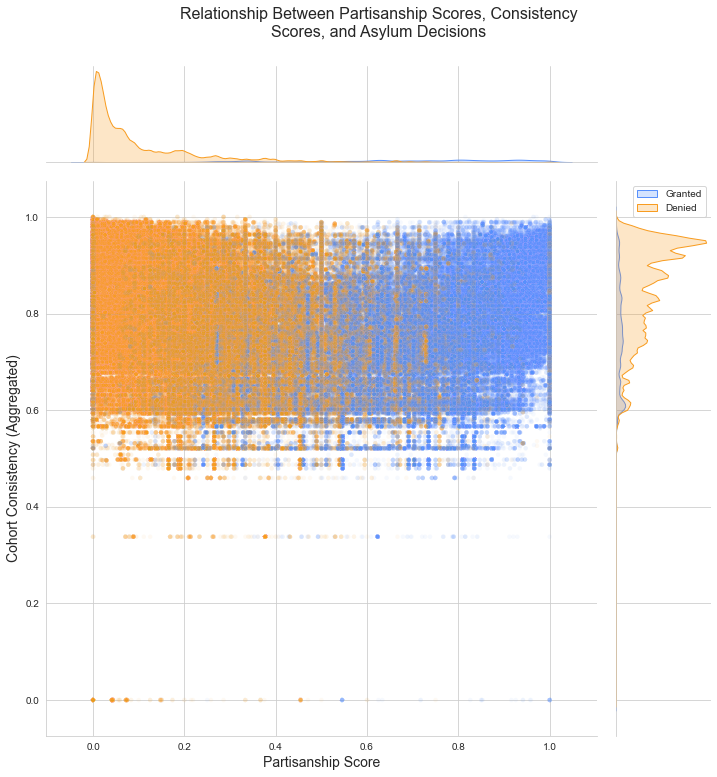}
\centering
\label{fig:agg_scores_decision}
\end{figure}

\begin{figure}
\caption{Covariance matrix for logistic regression built on partisanship and cohort consistency scores.}
\includegraphics[width=\linewidth]{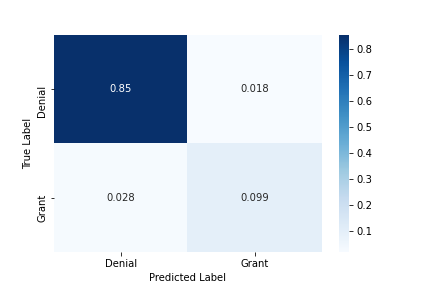}
\centering
\label{fig:svc_part}
\end{figure}

\subsubsection{Partisanship and Cohort Consistency} \label{Partisanship and Cohort Consistency}

Figure \ref{fig:agg_scores_decision} uses a jointplot to visualize asylum grants and denials across the different types of variability scores. Notably, grants seem to be concentrated in regions with high partisanship, while denials are concentrated in areas with low partisanship. This observation is in concordance with Figure \ref{fig:svc_part}, which illustrates average partisanship score and grant rate. In contrast, cohort consistency is not visually associated with either decision type. The standardized covariance between partisanship and cohort consistency, moreover, is -0.010823, suggesting that they are uncorrelated or only mildly so.

We probed these observations further using a logistic regression to classify grants and denials. We trained the model on 4,012,194 data points for which both cohort consistency scores and partisanship scores could be calculated. The accuracy of this model is 0.953477. While this initially seems quite high, both scores used to predict the decision were calculated using the decision; this results in leakage since researchers would not be able to implement the model in practice. Figure \ref{fig:svc_part} depicts the covariance matrix for the model. However, a few other statistics are noteworthy as well: the R$^2$ value is 0.582006, implying that these two scores cumulatively capture 58.2\% of the variability in the decision. Additionally, the weight of the partisanship score in this model is 10.035049, while the weight of the cohort consistency score is -4.519468. These numbers suggest that a high partisanship score and a low cohort consistency score imply a grant.

Observing the shape of the visualization, we hypothesized that a linear decision boundary may be sufficient for decision prediction using the variability scores. We therefore trained an additional model using a linear support vector classifier. This model had slightly higher performance, with an accuracy of 0.95386 and an R$^2$ of 0.585447. The partisanship weight of the decision boundary is 2.881612, while the cohort consistency weight is -0.893677. This replicates the general associations found by our logistic regression.

\paragraph{Partisanship and Disaggregated Consistency}
To investigate these results further, we also chose to examine the effect of disaggregated consistency scores. These values represent judge agreement for particular decisions in the step before averaging.

\begin{figure}
\caption{Decision distribution across partisanship and disaggregated consistency scores.}
\includegraphics[width=\linewidth]{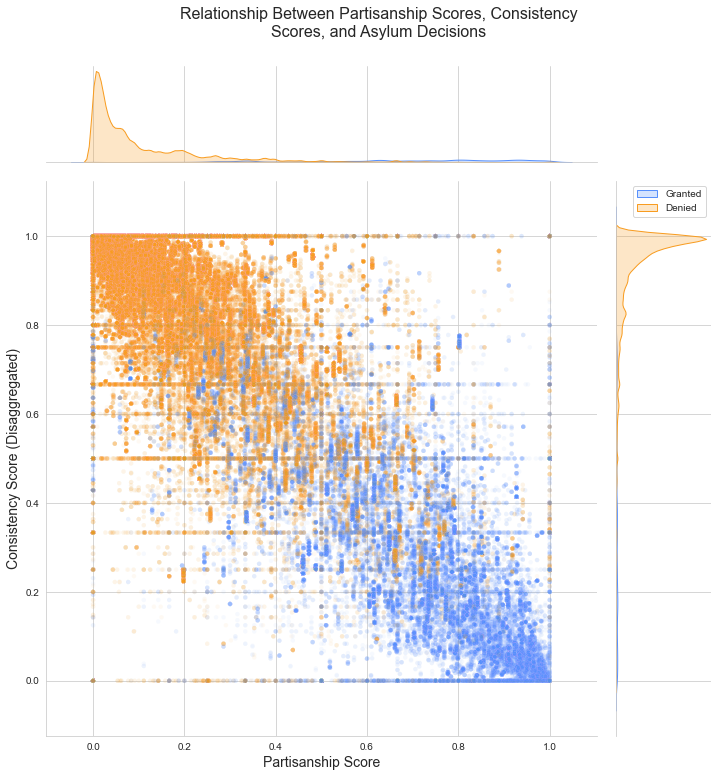}
\centering
\label{fig:disagg_scores_decision}
\end{figure}

Figure \ref{fig:disagg_scores_decision} depicts asylum grants and denials across partisanship scores and disaggregated consistency scores. This shows a more specific relationship with consistency: granted decisions are associated with lower disaggregated consistency. The standardized covariance between partisanship and disaggregated consistency, moreover, is -0.825603, suggesting opposing behavior of the scores.

We trained a logistic regression on this set of features on 817,666 data points. Its accuracy is slightly lower at 0.948304, and its R$^2$ value is 0.503281. The model weights are 4.493468 for partisanship and -6.333169 for disaggregated consistency. Compared to the models from the previous feature set, the consistency score has a much greater influence on the ultimate decision.

\subsubsection{Bivariate Prediction}
To further discern the role of different types of variability, we performed a similar analysis using partisanship, cohort consistency, and disaggregated consistency as the sole explanatory variables.

\paragraph{Partisanship}
First, we trained a logistic regression on the data points that had partisanship scores. This model has a 0.954842 accuracy and a 0.592405 R$^2$. Both of these performance statistics are higher on the partisanship-only model than on the ones that combine partisanship with consistency. This suggests that the contribution of consistency scores to the model may in fact detract from the accuracy of our model. We also computed the Spearman coefficient for the relationship between partisanship and decision to be 0.548642 (p = 0.0).

\begin{figure}
\caption{Covariance matrix for logistic regression built on cohort consistency scores.}
\includegraphics[width=\linewidth]{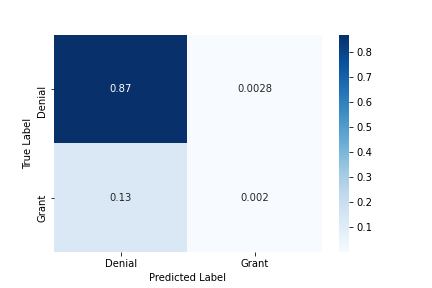}
\centering
\label{fig:logreg_cohort}
\end{figure}

\paragraph{Cohort Consistency}
To test our hypothesis about the consistency scores, we performed the same computation for both. The logistic regression for cohort consistency has an accuracy of 0.869808 and an R$^2$ of -0.155715. These statistics indicate that the model performs worse than the baseline classifier that always predicts asylum claim denial. The precision of this model is 0.416094 and the recall is 0.015307, suggesting that cohort consistency provides slightly more knowledge of positive asylum decisions than the baseline even as it results in lower accuracy. These findings are supported by the confusion matrix in Figure \ref{fig:logreg_cohort}. The Spearman correlation between cohort consistency and decision is -0.270801 (p = 0.0), indicating a mild negative relationship.

\paragraph{Disaggregated Consistency}
On the other hand, the logistic regression for disaggregated consistency has similar performance metrics to the other models. Its accuracy is 0.95043 and R$^2$ is 0.52186, and the precision and recall are 0.829881 and 0.727072 respectively. The Spearman correlation coefficient is -0.527486 (p = 0.0). The discrepancy between disaggregated and cohort (aggregated) consistency has implications for judicial behavior that are further explored in the following section.

\section{Discussion}

\subsection{Quantifying Decision Variability}
The goal of this study was to quantitatively discern the relative effects of systemic factors on asylum decisions, operationalized through our partisanship score, and individual variability, represented by our cohort consistency score. As described in Section \ref{Partisanship and Cohort Consistency}, we answered this question by constructing a logistic regression and a linear support vector classifier on partisanship and aggregated consistency scores. Notably, these two scores alone captured 58.5447\% of all case variability.

This statistically significant finding is troubling because it confirms that variables designed to represent bias are highly predictive a case’s outcome. Since the political climate and the assigned immigration judge are not within the applicant’s control and are often unrelated to the fact pattern of a case, these findings show a divergence from the U.S. immigration system’s arguable obligation to provide due process for all, even as due process rights for asylum applicants remain controversial since the U.S. Supreme Court has repeatedly circumvented the constitutional question to make rulings on “purely statutory grounds” \cite{due_process_2021, kendall_coffey_due_2001}.

Additionally, our results show that asylum grants are associated with high partisanship, while  denials are associated with low partisanship. This finding reflects the tendency of immigration courts to deny asylum (87.125\%); variability measured by the partisanship score occurs when immigration judges deviate from this trend by granting asylum.

\subsection{Consistency Aggregation}
The models' performances differed based on whether the consistency scores were aggregated or disaggregated. The aggregated models use cohort consistency scores, which are averaged across all cases for each judge, for prediction, while the disaggregated models use the consistency score for each individual case. We found that disaggregated consistency scores are more indicative of individual asylum case decisions than aggregated consistency scores.

This discrepancy gives the appearance that our model loses information through the aggregation process. Almost all aggregated (averaged) cohort consistency scores cluster above a 0.5 cohort consistency score. In contrast, disaggregated (individual case) consistency scores range from 0 to 1. This may indicate that judges’ decisions across all their cases are fairly consistent and that individual cases with low consistency scores are anomalous compared to the general behavior of the judge.

\subsection{Change in Partisanship Scores over Time}
Our time series analysis shows that partisanship in asylum decisions has increased over time, and particularly since the year 1992. Recall that in Section \ref{Systemic Variability: Partisanship} we have quantitatively defined partisanship as a measure of how much a decision varies across political climates. The increased partisanship trend in asylum decisions reflects increased political polarization nationally and populism globally \cite{drutman_american_2016}. The rise in partisanship can be attributed to several factors, including the susceptibility of voters to extreme narratives and the influence of non-civilian actors \cite{huddy_political_2017}. Simultaneously, the growth of populism creates suspicion towards policies like open immigration and globalization, prompting politicians to campaign on these divisive issues \cite{editors_of_the_world_politics_review_whats_2020, besley_rise_2021}.

Our time series analysis also indicates that the rise in partisanship slowed and then decreased around 2002. This finding may represent the rise in national unity (and thus decrease in partisanship) following the terrorist attacks on September 11, 2001 \cite{dan_balz_after_2021}.  However, partisanship metrics began to rise again beginning in 2014, and have reached their historical peak in 2022; this is evocative of the current polarization in national politics \cite{pew_research_center_political_2014}.

\subsection{Women as Outliers}
Our analysis found that the consistency scores of female judges were significantly lower than those of male judges. In the context of asylum denials being the norm, these findings confirm those in prior literature that state that women have higher asylum grant rates. This gender differential that labels women judges as “different” from the mean is notable considering that women were only slightly less represented than men among the judges in our dataset.\footnote{Our dataset featured 234 female judges compared to 365 male judges} Artificial intelligence and social science research have begun to consider the implications of characterizing women and minorities as statistical anomalies \cite{welles_minorities_2014}. Quantitative analysis may conflate statistical minorities with societal minorities by disproportionately flagging instances of a minority group as an outlier, which can lead to discriminatory findings \cite{shekhar_fairod_2021}.

\paragraph{Dataset Limitations} Due to the method of extracting biographical data about immigration judges, our dataset infers a judge’s gender from pronouns. This leads to a binary notion of gender as either male or female.

\subsection{Future Action}
We were struck by the extent of null data within the Executive Office for Immigration Review’s (EOIR) datasets, which indicates that much of the data is missing. Based on this observation, we join the Transactional Records Access Clearinghouse in calling for the EOIR to engage in more consistent and reliable data recording and dissemination processes \cite{trac_new_data_after_2020}. We also recommend that the EOIR statistically analyze specific judge demographics with low cohort consistencies to identify if there are “outlier” immigration judges within that sample that are ruling differently than their peers. By flagging individuals within specific “cohorts,” our research would enable the Department of Justice to take concrete steps towards further training and supervision to regulate the application of asylum law and ensure that due process and equal treatment are being met. Our partisanship metrics could likewise be used to monitor the health of the immigration court system longitudinally.

Following studies could integrate country-specific analyses, given the targeted treatment towards individuals coming from certain countries and the impact of the quota system on court-by-court decisions.\footnote{For example, see the targeted treatment and deportations of Haitians without trial under Title 42 and the pretext of public health \cite{sullivan_biden_2021}.} Our methodology can be adapted from the immigration court system to the federal and state court systems, where the hierarchical nature of \textit{stare decisis} provides further justification for the importance of inter-court consistency. Future research could also aggregate consistency scores by immigration court rather than by judge to focus on intra-court inconsistencies rather than inter-court inconsistencies as we have done.

\subsection{Conclusion}
In this paper, we introduced a novel scoring device that accurately captures partisanship and cohort consistency metrics in order to close a critical gap in the literature of quantification of asylum decision variability that can be compared across spatial and temporal boundaries. Using predictive modeling, we found that immigration court decisions are predominantly impacted by two extraneous factors: the surrounding political climate and the individual variability of the presiding judge. Our contributions leverage computing as a diagnostic tool by attempting to improve and standardize the measurement of variability in the U.S. immigration court system \cite{abebe_roles_2020}. Ultimately, we seek to uplift equity and justice in asylum adjudication by increasing accountability for these powerful systems.

\begin{acks}
We are indebted to the following researchers at the Human Rights Center Investigations Lab at the University of California, Berkeley, whose efforts have been integral to the success of this project: Aarushi Karandikar, Aayush Patel, Aliyah Behimino, Carolyn Wang, Eleanor Wong, Elena Wüllhorst, Helia Sadeghi, Karina Cortes Garcia, Margaret Carroll, Rosie Foulds, Upasana Dilip. We would also like to thank our advisors at the Human Rights Center who have offered guidance throughout this process: Dr. Alexa Koenig, Stephanie Croft, Sofia-Lisset Kooner.

Finally, we would like to acknowledge the following individuals for their generous review of our work: Connor Haley, consultant, UC Berkeley D-Lab; Lisa Knox, Legal Director, California Collaborative for Immigrant Justice; Dr. Brandie Nonnecke, Director, CITRIS Policy Lab; Kate Jastram, Director of Policy \& Advocacy at the Center for Gender \& Refugee Studies, UC Hastings College of the Law; Ezinne Nwankwo, PhD Student, UC Berkeley Electrical Engineering and Computer Science; Emma Lurie, PhD Student, UC Berkeley School of Information.
\end{acks}

\bibliographystyle{ACM-Reference-Format}
\bibliography{bias_cons_part_asylum.bib}


\begin{thebibliography}{48}


\ifx \showCODEN    \undefined \def \showCODEN     #1{\unskip}     \fi
\ifx \showDOI      \undefined \def \showDOI       #1{#1}\fi
\ifx \showISBNx    \undefined \def \showISBNx     #1{\unskip}     \fi
\ifx \showISBNxiii \undefined \def \showISBNxiii  #1{\unskip}     \fi
\ifx \showISSN     \undefined \def \showISSN      #1{\unskip}     \fi
\ifx \showLCCN     \undefined \def \showLCCN      #1{\unskip}     \fi
\ifx \shownote     \undefined \def \shownote      #1{#1}          \fi
\ifx \showarticletitle \undefined \def \showarticletitle #1{#1}   \fi
\ifx \showURL      \undefined \def \showURL       {\relax}        \fi
\providecommand\bibfield[2]{#2}
\providecommand\bibinfo[2]{#2}
\providecommand\natexlab[1]{#1}
\providecommand\showeprint[2][]{arXiv:#2}

\bibitem[imm(1952)]%
        {immigration_nationality_act}
 \bibinfo{year}{1952}\natexlab{}.
\newblock \bibinfo{title}{Immigration and {Nationality} {Act}}.
\newblock
\newblock
\urldef\tempurl%
\url{https://www.uscis.gov/laws-and-policy/legislation/immigration-and-nationality-act}
\showURL{%
\tempurl}


\bibitem[con(1967)]%
        {convention_refugees}
 \bibinfo{year}{1967}\natexlab{}.
\newblock \bibinfo{title}{Convention and {Protocol} {Relating} to the {Status}
  of {Refugees}}.
\newblock
\newblock


\bibitem[pro(2011)]%
        {professionalism_guide}
 \bibinfo{year}{2011}\natexlab{}.
\newblock \bibinfo{title}{Ethics and {Professionalism} {Guide} for
  {Immigration} {Judges}}.
\newblock
\newblock
\urldef\tempurl%
\url{https://www.justice.gov/sites/default/files/eoir/legacy/2013/05/23/EthicsandProfessionalismGuideforIJs.pdf}
\showURL{%
\tempurl}


\bibitem[tra(2020)]%
        {trac_new_data_after_2020}
 \bibinfo{year}{2020}\natexlab{}.
\newblock \bibinfo{title}{After {EOIR} {Fixes} {Most} {Egregious} {Data}
  {Errors}, {TRAC} {Releases} {New} {Asylum} {Data} - {But} with a {Warning}}.
\newblock
\newblock
\urldef\tempurl%
\url{https://trac.syr.edu/immigration/reports/624}
\showURL{%
\tempurl}


\bibitem[bia(2021)]%
        {bia_board}
 \bibinfo{year}{2021}\natexlab{}.
\newblock \bibinfo{title}{Board of {Immigration} {Appeals}}.
\newblock
\newblock
\urldef\tempurl%
\url{https://www.justice.gov/eoir/board-of-immigration-appeals}
\showURL{%
\tempurl}


\bibitem[due(2021)]%
        {due_process_2021}
 \bibinfo{year}{2021}\natexlab{}.
\newblock \bibinfo{title}{Due {Process} in {Immigration} {Proceedings}}.
\newblock
\newblock
\urldef\tempurl%
\url{https://cdn.ca9.uscourts.gov/datastore/uploads/immigration/immig_west/E.pdf}
\showURL{%
\tempurl}


\bibitem[aff(2022)]%
        {affirmative}
 \bibinfo{year}{2022}\natexlab{}.
\newblock \bibinfo{title}{The {Affirmative} {Asylum} {Process}}.
\newblock
\newblock
\urldef\tempurl%
\url{https://www.uscis.gov/humanitarian/refugees-and-asylum/asylum/the-affirmative-asylum-process}
\showURL{%
\tempurl}


\bibitem[tra(2022)]%
        {trac_tool}
 \bibinfo{year}{2022}\natexlab{}.
\newblock \bibinfo{title}{Asylum Decision Tool}.
\newblock
\newblock
\urldef\tempurl%
\url{https://trac.syr.edu/phptools/immigration/asylum}
\showURL{%
\tempurl}


\bibitem[typ(2022)]%
        {typesOfAsylum}
 \bibinfo{year}{2022}\natexlab{}.
\newblock \bibinfo{title}{Types of {Asylum}}.
\newblock
\newblock
\urldef\tempurl%
\url{https://help.unhcr.org/usa/applying-for-asylum/types-of-asylum}
\showURL{%
\tempurl}


\bibitem[Abebe et~al\mbox{.}(2020)]%
        {abebe_roles_2020}
\bibfield{author}{\bibinfo{person}{Rediet Abebe}, \bibinfo{person}{Solon
  Barocas}, \bibinfo{person}{Jon Kleinberg}, \bibinfo{person}{Karen Levy},
  \bibinfo{person}{Manish Raghavan}, {and} \bibinfo{person}{David~G.
  Robinson}.} \bibinfo{year}{2020}\natexlab{}.
\newblock \showarticletitle{Roles for Computing in Social Change}. In
  \bibinfo{booktitle}{\emph{Proceedings of the 2020 Conference on Fairness,
  Accountability, and Transparency}} (Barcelona, Spain)
  \emph{(\bibinfo{series}{FAT* '20})}. \bibinfo{publisher}{Association for
  Computing Machinery}, \bibinfo{address}{New York, NY, USA},
  \bibinfo{pages}{252–260}.
\newblock
\showISBNx{9781450369367}
\urldef\tempurl%
\url{https://doi.org/10.1145/3351095.3372871}
\showDOI{\tempurl}


\bibitem[Ash et~al\mbox{.}(2007)]%
        {chen_machine_predict}
\bibfield{author}{\bibinfo{person}{Elliott Ash}, \bibinfo{person}{Daniel Chen},
  \bibinfo{person}{Colin Andrus}, \bibinfo{person}{Dustin Godevais}, {and}
  \bibinfo{person}{Gary Ng}.} \bibinfo{year}{2007}\natexlab{}.
\newblock \showarticletitle{Machine {Prediction} of {Appeal} {Success} in
  {U}.{S}. {Asylum} {Courts}}.
\newblock  (\bibinfo{year}{2007}).
\newblock
\urldef\tempurl%
\url{https://users.nber.org/~dlchen/papers/Machine_Prediction_of_Appeal_Success_in_US_Asylum_Courts.pdf}
\showURL{%
\tempurl}


\bibitem[Besley and Persson(2021)]%
        {besley_rise_2021}
\bibfield{author}{\bibinfo{person}{Timothy Besley} {and}
  \bibinfo{person}{Torsten Persson}.} \bibinfo{year}{2021}\natexlab{}.
\newblock \bibinfo{title}{The {Rise} of {Identity} {Politics}: {Policy},
  {Political} {Organization}, and {Nationalist} {Dynamics}}.
\newblock
\newblock
\urldef\tempurl%
\url{https://www.lse.ac.uk/economics/Assets/Documents/personal-pages/tim-besley/working-papers/the-rise-of-identity-politics.pdf}
\showURL{%
\tempurl}


\bibitem[Brace et~al\mbox{.}(2000)]%
        {brace_measuring_2000}
\bibfield{author}{\bibinfo{person}{Paul Brace}, \bibinfo{person}{Laura Langer},
  {and} \bibinfo{person}{Melinda~Gann Hall}.} \bibinfo{year}{2000}\natexlab{}.
\newblock \showarticletitle{Measuring the {Preferences} of {State} {Supreme}
  {Court} {Judges}}.
\newblock \bibinfo{journal}{\emph{The Journal of Politics}}
  \bibinfo{volume}{62}, \bibinfo{number}{2} (\bibinfo{date}{May}
  \bibinfo{year}{2000}), \bibinfo{pages}{387--413}.
\newblock
\showISSN{0022-3816, 1468-2508}
\urldef\tempurl%
\url{https://doi.org/10.1111/0022-3816.00018}
\showDOI{\tempurl}
\newblock
\shownote{Number: 2}.


\bibitem[Breiman(2001)]%
        {breiman_statistical_2001}
\bibfield{author}{\bibinfo{person}{Leo Breiman}.}
  \bibinfo{year}{2001}\natexlab{}.
\newblock \showarticletitle{Statistical {Modeling}: {The} {Two} {Cultures}}.
\newblock \bibinfo{journal}{\emph{Statist. Sci.}} \bibinfo{volume}{16},
  \bibinfo{number}{3} (\bibinfo{year}{2001}), \bibinfo{pages}{199 -- 231}.
\newblock
\urldef\tempurl%
\url{https://doi.org/10.1214/ss/1009213726}
\showDOI{\tempurl}
\newblock
\shownote{Number: 3 Publisher: Institute of Mathematical Statistics}.


\bibitem[Chen and Eagel(2017a)]%
        {chen_proceedings_2017}
\bibfield{author}{\bibinfo{person}{Daniel Chen} {and} \bibinfo{person}{Jess
  Eagel}.} \bibinfo{year}{2017}\natexlab{a}.
\newblock \showarticletitle{Proceedings of the 16th edition of the
  {International} {Conference} on {Articial} {Intelligence} and {Law}}.
  \bibinfo{publisher}{Association for Computing Machinery}.
\newblock


\bibitem[Chen et~al\mbox{.}(2016)]%
        {chen_decision-making}
\bibfield{author}{\bibinfo{person}{Daniel Chen}, \bibinfo{person}{Tobias~J
  Moskowitz}, {and} \bibinfo{person}{Kelly Shue}.}
  \bibinfo{year}{2016}\natexlab{}.
\newblock \showarticletitle{Decision-{Making} under the {Gambler}'s {Fallacy}:
  {Evidence} from {Asylum} {Judges}, {Loan} {Officers}, and {Baseball}
  {Umpires}}.
\newblock \bibinfo{journal}{\emph{The Quarterly Journal of Economics}}
  \bibinfo{volume}{131}, \bibinfo{number}{3} (\bibinfo{date}{March}
  \bibinfo{year}{2016}), \bibinfo{pages}{1181--1242}.
\newblock
\urldef\tempurl%
\url{https://doi.org/10.1093/qje/qjw017}
\showDOI{\tempurl}


\bibitem[Chen and Eagel(2017b)]%
        {chenCan2017}
\bibfield{author}{\bibinfo{person}{Daniel~L. Chen} {and} \bibinfo{person}{Jess
  Eagel}.} \bibinfo{year}{2017}\natexlab{b}.
\newblock \showarticletitle{Can machine learning help predict the outcome of
  asylum adjudications?}. In \bibinfo{booktitle}{\emph{Proceedings of the 16th
  edition of the {International} {Conference} on {Articial} {Intelligence} and
  {Law}}}. \bibinfo{publisher}{ACM}, \bibinfo{address}{London United Kingdom},
  \bibinfo{pages}{237--240}.
\newblock
\showISBNx{978-1-4503-4891-1}
\urldef\tempurl%
\url{https://doi.org/10.1145/3086512.3086538.}
\showDOI{\tempurl}


\bibitem[Cropanzano et~al\mbox{.}(2015)]%
        {cropanzano_measuring_2015}
\bibfield{author}{\bibinfo{person}{Russell~S. Cropanzano},
  \bibinfo{person}{Maureen~L. Ambrose}, \bibinfo{person}{Jason~A. Colquitt},
  {and} \bibinfo{person}{Jessica~B. Rodell}.} \bibinfo{year}{2015}\natexlab{}.
\newblock \showarticletitle{Measuring {Justice} and {Fairness}}.
\newblock In \bibinfo{booktitle}{\emph{The {Oxford} {Handbook} of {Justice} in
  the {Workplace}}}, \bibfield{editor}{\bibinfo{person}{Russell~S. Cropanzano}
  {and} \bibinfo{person}{Maureen~L. Ambrose}} (Eds.).
  \bibinfo{publisher}{Oxford University Press}.
\newblock
\showISBNx{978-0-19-998141-0}
\urldef\tempurl%
\url{https://doi.org/10.1093/oxfordhb/9780199981410.013.8}
\showDOI{\tempurl}


\bibitem[{Dan Balz}(2021)]%
        {dan_balz_after_2021}
\bibfield{author}{\bibinfo{person}{{Dan Balz}}.}
  \bibinfo{year}{2021}\natexlab{}.
\newblock \bibinfo{title}{After 9/11, a rush of national unity. {Then},
  quickly, more and new divisions}.
\newblock
\newblock
\urldef\tempurl%
\url{https://www.washingtonpost.com/politics/after-911-a-rush-of-national-unity-then-quickly-more-and-new-divisions/2021/09/11/8f6f7d8e-12a9-11ec-bc8a-8d9a5b534194_story.html}
\showURL{%
\tempurl}


\bibitem[Danziger et~al\mbox{.}(2011)]%
        {danziger_extraneous_2011}
\bibfield{author}{\bibinfo{person}{Shai Danziger}, \bibinfo{person}{Jonathan
  Levav}, {and} \bibinfo{person}{Liora Avnaim-Pesso}.}
  \bibinfo{year}{2011}\natexlab{}.
\newblock \showarticletitle{Extraneous {Factors} in {Judicial} {Decisions}}.
\newblock \bibinfo{journal}{\emph{Proceedings of the National Academy of
  Sciences}} \bibinfo{volume}{108}, \bibinfo{number}{17} (\bibinfo{date}{April}
  \bibinfo{year}{2011}), \bibinfo{pages}{6889--6892}.
\newblock
\urldef\tempurl%
\url{https://doi.org/10.1073/pnas.1018033108}
\showDOI{\tempurl}
\newblock
\shownote{Number: 17 Publisher: Proceedings of the National Academy of
  Sciences}.


\bibitem[Drutman(2016)]%
        {drutman_american_2016}
\bibfield{author}{\bibinfo{person}{Lee Drutman}.}
  \bibinfo{year}{2016}\natexlab{}.
\newblock \showarticletitle{American politics has reached peak polarization}.
\newblock \bibinfo{journal}{\emph{Vox}} (\bibinfo{date}{March}
  \bibinfo{year}{2016}).
\newblock
\urldef\tempurl%
\url{https://www.vox.com/polyarchy/2016/3/24/11298808/american-politics-peak-polarization}
\showURL{%
\tempurl}


\bibitem[Dunn et~al\mbox{.}(2017)]%
        {dunn_early_2017}
\bibfield{author}{\bibinfo{person}{Matt Dunn}, \bibinfo{person}{Levent Sagun},
  \bibinfo{person}{Hale Şirin}, {and} \bibinfo{person}{Daniel Chen}.}
  \bibinfo{year}{2017}\natexlab{}.
\newblock \showarticletitle{Early predictability of asylum court decisions}. In
  \bibinfo{booktitle}{\emph{Proceedings of the 16th edition of the
  {International} {Conference} on {Articial} {Intelligence} and {Law}}}.
  \bibinfo{publisher}{ACM}, \bibinfo{address}{London United Kingdom},
  \bibinfo{pages}{233--236}.
\newblock
\showISBNx{978-1-4503-4891-1}
\urldef\tempurl%
\url{https://doi.org/10.1145/3086512.3086537.}
\showDOI{\tempurl}


\bibitem[{Editors of the World Politics Review.}(2020)]%
        {editors_of_the_world_politics_review_whats_2020}
\bibfield{author}{\bibinfo{person}{{Editors of the World Politics Review.}}}
  \bibinfo{year}{2020}\natexlab{}.
\newblock \bibinfo{title}{What’s {Driving} the {Rise} of {Authoritarianism}
  and {Populism} in {Europe} and {Beyond}?}
\newblock
\newblock
\urldef\tempurl%
\url{https://www.worldpoliticsreview.com/insights/27842/the-rise-of-authoritarianism-and-populism-europe-and-beyond}
\showURL{%
\tempurl}


\bibitem[Hu et~al\mbox{.}(2019)]%
        {hu_auditing_2019}
\bibfield{author}{\bibinfo{person}{Desheng Hu}, \bibinfo{person}{Shan Jiang},
  \bibinfo{person}{Ronald E.~Robertson}, {and} \bibinfo{person}{Christo
  Wilson}.} \bibinfo{year}{2019}\natexlab{}.
\newblock \showarticletitle{Auditing the {Partisanship} of {Google} {Search}
  {Snippets}}. In \bibinfo{booktitle}{\emph{The {World} {Wide} {Web}
  {Conference} on - {WWW} '19}}. \bibinfo{publisher}{ACM Press},
  \bibinfo{address}{San Francisco, CA, USA}, \bibinfo{pages}{693--704}.
\newblock
\showISBNx{978-1-4503-6674-8}
\urldef\tempurl%
\url{https://doi.org/10.1145/3308558.3313654.}
\showDOI{\tempurl}


\bibitem[Huddy and Bankert(2017)]%
        {huddy_political_2017}
\bibfield{author}{\bibinfo{person}{Leonie Huddy} {and} \bibinfo{person}{Alexa
  Bankert}.} \bibinfo{year}{2017}\natexlab{}.
\newblock \showarticletitle{Political {Partisanship} as a {Social} {Identity}}.
\newblock In \bibinfo{booktitle}{\emph{Oxford {Research} {Encyclopedia} of
  {Politics}}}. \bibinfo{publisher}{Oxford University Press}.
\newblock
\showISBNx{978-0-19-022863-7}
\urldef\tempurl%
\url{https://doi.org/10.1093/acrefore/9780190228637.013.250.}
\showDOI{\tempurl}


\bibitem[Hunter(2007)]%
        {hunter_matplotlib_2007}
\bibfield{author}{\bibinfo{person}{John~D. Hunter}.}
  \bibinfo{year}{2007}\natexlab{}.
\newblock \showarticletitle{Matplotlib: {A} {2D} {Graphics} {Environment}}.
\newblock \bibinfo{journal}{\emph{Computing in Science \& Engineering}}
  \bibinfo{volume}{9}, \bibinfo{number}{3} (\bibinfo{year}{2007}),
  \bibinfo{pages}{90--95}.
\newblock
\showISSN{1521-9615}
\urldef\tempurl%
\url{https://doi.org/10.1109/MCSE.2007.55.}
\showDOI{\tempurl}


\bibitem[Ippolito and Velluti(2011)]%
        {ippolito_recast_2011}
\bibfield{author}{\bibinfo{person}{Francesca Ippolito} {and}
  \bibinfo{person}{Samantha Velluti}.} \bibinfo{year}{2011}\natexlab{}.
\newblock \showarticletitle{{The} {Recast} {Process} {Of} {The} {EU} {Asylum}
  {System}: {A} {Balancing} {Act} {Between} {Efficiency} {And} {Fairness}}.
\newblock \bibinfo{journal}{\emph{Refugee Survey Quarterly}}
  \bibinfo{volume}{30}, \bibinfo{number}{3} (\bibinfo{year}{2011}),
  \bibinfo{pages}{24--62}.
\newblock


\bibitem[Johnson(2019)]%
        {johnson_isaiah_under_2019}
\bibfield{author}{\bibinfo{person}{Isaiah Johnson}.}
  \bibinfo{year}{2019}\natexlab{}.
\newblock \showarticletitle{{Under} {Pressure}: {Measuring} {Constituent}
  {Attitudes} {On} {Immigration} {And} {Its} {Effects} {On} {Legislative}
  {Behavior}}.
\newblock  (\bibinfo{date}{May} \bibinfo{year}{2019}).
\newblock
\urldef\tempurl%
\url{https://uh-ir.tdl.org/bitstream/handle/10657/4261/Johnson_Isaiah_2019URD.pdf?sequence=1&isAllowed=y}
\showURL{%
\tempurl}


\bibitem[Kamalaharan and Gutman(2017)]%
        {kamalaharan_predicting}
\bibfield{author}{\bibinfo{person}{Dagshayani Kamalaharan} {and}
  \bibinfo{person}{Jacqueline Gutman}.} \bibinfo{year}{2017}\natexlab{}.
\newblock \showarticletitle{Predicting {Asylum} {Court} {Decisions} and
  {Detecting} {Outlier} {Judges}}.
\newblock  (\bibinfo{date}{Oct.} \bibinfo{year}{2017}), \bibinfo{pages}{23}.
\newblock


\bibitem[{Kendall Coffey}(2001)]%
        {kendall_coffey_due_2001}
\bibfield{author}{\bibinfo{person}{{Kendall Coffey}}.}
  \bibinfo{year}{2001}\natexlab{}.
\newblock \showarticletitle{The {Due} {Process} {Right} to {Seek} {Asylum} in
  the {United} {States}: {The} {Immigration} {Dilemma} and {Constitutional}
  {Controversy}.}
\newblock \bibinfo{journal}{\emph{The Yale Law and Policy Review}}
  \bibinfo{volume}{19} (\bibinfo{year}{2001}), \bibinfo{pages}{303--339}.
\newblock
\urldef\tempurl%
\url{https://www.jstor.org/stable/40239567}
\showURL{%
\tempurl}


\bibitem[Kim and Semet(2020)]%
        {kim_empirical}
\bibfield{author}{\bibinfo{person}{Catherine Kim} {and} \bibinfo{person}{Amy
  Semet}.} \bibinfo{year}{2020}\natexlab{}.
\newblock \showarticletitle{An {Empirical} {Study} of {Political} {Control}
  over {Immigration} {Adjudication}}.
\newblock \bibinfo{journal}{\emph{The Georgetown Law Journal}}
  \bibinfo{volume}{108}, \bibinfo{number}{579} (\bibinfo{date}{March}
  \bibinfo{year}{2020}), \bibinfo{pages}{579--467}.
\newblock
\urldef\tempurl%
\url{https://www.law.georgetown.edu/georgetown-law-journal/in-print/volume-108/volume-108-issue-3-march-2020/an-empirical-study-of-political-control-over-immigration-adjudication}
\showURL{%
\tempurl}


\bibitem[Kusner et~al\mbox{.}(2018)]%
        {kusner_counterfactual_2018}
\bibfield{author}{\bibinfo{person}{Matt~J. Kusner}, \bibinfo{person}{Joshua~R.
  Loftus}, \bibinfo{person}{Chris Russell}, {and} \bibinfo{person}{Ricardo
  Silva}.} \bibinfo{year}{2018}\natexlab{}.
\newblock \showarticletitle{Counterfactual {Fairness}}.
\newblock \bibinfo{journal}{\emph{arXiv:1703.06856 [cs, stat]}}
  (\bibinfo{date}{March} \bibinfo{year}{2018}).
\newblock
\urldef\tempurl%
\url{https://doi.org/10.48550/arXiv.1703.06856}
\showDOI{\tempurl}
\newblock
\shownote{arXiv: 1703.06856}.


\bibitem[Mehrabi et~al\mbox{.}(2022)]%
        {mehrabi_survey_2022}
\bibfield{author}{\bibinfo{person}{Ninareh Mehrabi}, \bibinfo{person}{Fred
  Morstatter}, \bibinfo{person}{Nripsuta Saxena}, \bibinfo{person}{Kristina
  Lerman}, {and} \bibinfo{person}{Aram Galstyan}.}
  \bibinfo{year}{2022}\natexlab{}.
\newblock \showarticletitle{A {Survey} on {Bias} and {Fairness} in {Machine}
  {Learning}}.
\newblock \bibinfo{journal}{\emph{arXiv:1908.09635 [cs]}} (\bibinfo{date}{Jan.}
  \bibinfo{year}{2022}).
\newblock
\urldef\tempurl%
\url{http://arxiv.org/abs/1908.09635}
\showURL{%
\tempurl}
\newblock
\shownote{arXiv: 1908.09635}.


\bibitem[pandas~development team(2020)]%
        {reback2020pandas}
\bibfield{author}{\bibinfo{person}{The pandas~development team}.}
  \bibinfo{year}{2020}\natexlab{}.
\newblock \bibinfo{booktitle}{\emph{pandas-dev/pandas: Pandas}}.
\newblock
\urldef\tempurl%
\url{https://doi.org/10.5281/zenodo.3509134}
\showDOI{\tempurl}


\bibitem[Pedregosa et~al\mbox{.}(2011)]%
        {scikit-learn}
\bibfield{author}{\bibinfo{person}{F. Pedregosa}, \bibinfo{person}{G.
  Varoquaux}, \bibinfo{person}{A. Gramfort}, \bibinfo{person}{V. Michel},
  \bibinfo{person}{B. Thirion}, \bibinfo{person}{O. Grisel},
  \bibinfo{person}{M. Blondel}, \bibinfo{person}{P. Prettenhofer},
  \bibinfo{person}{R. Weiss}, \bibinfo{person}{V. Dubourg}, \bibinfo{person}{J.
  Vanderplas}, \bibinfo{person}{A. Passos}, \bibinfo{person}{D. Cournapeau},
  \bibinfo{person}{M. Brucher}, \bibinfo{person}{M. Perrot}, {and}
  \bibinfo{person}{E. Duchesnay}.} \bibinfo{year}{2011}\natexlab{}.
\newblock \showarticletitle{Scikit-learn: Machine Learning in {P}ython}.
\newblock \bibinfo{journal}{\emph{Journal of Machine Learning Research}}
  \bibinfo{volume}{12} (\bibinfo{year}{2011}), \bibinfo{pages}{2825--2830}.
\newblock


\bibitem[{Pew Research Center}(2014)]%
        {pew_research_center_political_2014}
\bibfield{author}{\bibinfo{person}{{Pew Research Center}}.}
  \bibinfo{year}{2014}\natexlab{}.
\newblock \bibinfo{title}{Political {Polarization} in the {American} {Public}:
  {How} {Increasing} {Ideological} {Uniformity} and {Partisan} {Antipathy}
  {Affect} {Politics}, {Compromise} and {Everyday} {Life}.}
\newblock
\newblock
\urldef\tempurl%
\url{https://www.pewresearch.org/politics/2014/06/12/political-polarization-in-the-american-public}
\showURL{%
\tempurl}


\bibitem[{Reuters}(2017)]%
        {reuters_not-quite-independent_2017}
\bibfield{author}{\bibinfo{person}{{Reuters}}.}
  \bibinfo{year}{2017}\natexlab{}.
\newblock \showarticletitle{The {Not}-{Quite}-{Independent} {U}.{S}.
  {Immigration} {Courts}}.
\newblock  (\bibinfo{date}{Oct.} \bibinfo{year}{2017}).
\newblock
\urldef\tempurl%
\url{https://www.reuters.com/article/uk-usa-immigration-court-idUKKBN1CM1UO}
\showURL{%
\tempurl}


\bibitem[Salehyan and Rosenblum(2008)]%
        {salehyan_international_2008}
\bibfield{author}{\bibinfo{person}{Idean Salehyan} {and} \bibinfo{person}{Marc
  Rosenblum}.} \bibinfo{year}{2008}\natexlab{}.
\newblock \showarticletitle{International {Relations}, {Domestic} {Politics},
  and {Asylum} {Admissions} in the {United} {States}}.
\newblock \bibinfo{journal}{\emph{Political Research Quarterly}}
  \bibinfo{volume}{61}, \bibinfo{number}{1} (\bibinfo{year}{2008}),
  \bibinfo{pages}{104--121}.
\newblock


\bibitem[Schoenholtz et~al\mbox{.}(2009)]%
        {refugeeRoulette}
\bibfield{author}{\bibinfo{person}{Andrew~I. Schoenholtz},
  \bibinfo{person}{Jaya Ramji-Nogales}, {and} \bibinfo{person}{Philip~G.
  Schrag}.} \bibinfo{year}{2009}\natexlab{}.
\newblock \bibinfo{booktitle}{\emph{Refugee {Roulette}: {Disparities} in
  {Asylum} {Adjudication}}}.
\newblock \bibinfo{publisher}{NYU Press}.
\newblock


\bibitem[Shekhar et~al\mbox{.}(2021)]%
        {shekhar_fairod_2021}
\bibfield{author}{\bibinfo{person}{Shubhranshu Shekhar}, \bibinfo{person}{Neil
  Shah}, {and} \bibinfo{person}{Leman Akoglu}.}
  \bibinfo{year}{2021}\natexlab{}.
\newblock \showarticletitle{{FairOD}: {Fairness}-aware {Outlier} {Detection}}.
  In \bibinfo{booktitle}{\emph{Proceedings of the 2021 {AAAI}/{ACM}
  {Conference} on {AI}, {Ethics}, and {Society}}}. \bibinfo{publisher}{ACM},
  \bibinfo{address}{Virtual Event USA}, \bibinfo{pages}{210--220}.
\newblock
\showISBNx{978-1-4503-8473-5}
\urldef\tempurl%
\url{https://doi.org/10.1145/3461702.3462517}
\showDOI{\tempurl}


\bibitem[Sullivan and Jordan(2021)]%
        {sullivan_biden_2021}
\bibfield{author}{\bibinfo{person}{Eileen Sullivan} {and}
  \bibinfo{person}{Miriam Jordan}.} \bibinfo{year}{2021}\natexlab{}.
\newblock \showarticletitle{Biden {Administration} to {Deport} {Haitians} in
  {South} {Texas}}.
\newblock \bibinfo{journal}{\emph{The New York Times}} (\bibinfo{date}{Sept.}
  \bibinfo{year}{2021}).
\newblock
\urldef\tempurl%
\url{https://www.nytimes.com/2021/09/18/us/politics/biden-administration-haiti-texas.html}
\showURL{%
\tempurl}


\bibitem[Taylor and Letham(2017)]%
        {taylor_forecasting_2017}
\bibfield{author}{\bibinfo{person}{Sean~J Taylor} {and}
  \bibinfo{person}{Benjamin Letham}.} \bibinfo{year}{2017}\natexlab{}.
\newblock \bibinfo{booktitle}{\emph{Forecasting at {Scale}}}.
\newblock \bibinfo{type}{preprint}. \bibinfo{institution}{PeerJ Preprints}.
\newblock
\urldef\tempurl%
\url{https://doi.org/10.7287/peerj.preprints.3190v2.}
\showDOI{\tempurl}


\bibitem[Tyler(1984)]%
        {tyler_role_1984}
\bibfield{author}{\bibinfo{person}{T~R Tyler}.}
  \bibinfo{year}{1984}\natexlab{}.
\newblock \showarticletitle{Role of {Perceived} {Injustice} in {Defendants}'
  {Evaluations} of {Their} {Courtroom} {Experience}}.
\newblock \bibinfo{journal}{\emph{Law and Society Review}}
  \bibinfo{volume}{18}, \bibinfo{number}{1} (\bibinfo{year}{1984}),
  \bibinfo{pages}{51--74}.
\newblock
\urldef\tempurl%
\url{https://www.ojp.gov/ncjrs/virtual-library/abstracts/role-perceived-injustice-defendants-evaluations-their-courtroom}
\showURL{%
\tempurl}


\bibitem[{U.S. Government Accountability Office}(2016)]%
        {us_government_accountability_office_asylum_2016}
\bibfield{author}{\bibinfo{person}{{U.S. Government Accountability Office}}.}
  \bibinfo{year}{2016}\natexlab{}.
\newblock \showarticletitle{Asylum: {Variation} {Exists} in {Outcomes} of
  {Applications} {Across} {Immigration} {Courts} and {Judges}, {Accessible}
  {Version}}.
\newblock  (\bibinfo{date}{Nov.} \bibinfo{year}{2016}).
\newblock
\urldef\tempurl%
\url{https://www.gao.gov/assets/690/682965.pdf}
\showURL{%
\tempurl}


\bibitem[Waskom(2021)]%
        {waskom_seaborn_2021}
\bibfield{author}{\bibinfo{person}{Michael Waskom}.}
  \bibinfo{year}{2021}\natexlab{}.
\newblock \showarticletitle{seaborn: statistical data visualization}.
\newblock \bibinfo{journal}{\emph{Journal of Open Source Software}}
  \bibinfo{volume}{6}, \bibinfo{number}{60} (\bibinfo{date}{April}
  \bibinfo{year}{2021}), \bibinfo{pages}{3021}.
\newblock
\showISSN{2475-9066}
\urldef\tempurl%
\url{https://doi.org/10.21105/joss.03021.}
\showDOI{\tempurl}


\bibitem[Welles(2014)]%
        {welles_minorities_2014}
\bibfield{author}{\bibinfo{person}{Brooke~Foucault Welles}.}
  \bibinfo{year}{2014}\natexlab{}.
\newblock \showarticletitle{On minorities and outliers: {The} case for making
  {Big} {Data} small}.
\newblock \bibinfo{journal}{\emph{Big Data \& Society}} \bibinfo{volume}{1},
  \bibinfo{number}{1} (\bibinfo{date}{April} \bibinfo{year}{2014}),
  \bibinfo{pages}{205395171454061}.
\newblock
\showISSN{2053-9517, 2053-9517}
\urldef\tempurl%
\url{https://doi.org/10.1177/2053951714540613}
\showDOI{\tempurl}


\bibitem[{W}es {M}c{K}inney(2010)]%
        {mckinney-proc-scipy-2010}
\bibfield{author}{\bibinfo{person}{{W}es {M}c{K}inney}.}
  \bibinfo{year}{2010}\natexlab{}.
\newblock \showarticletitle{{D}ata {S}tructures for {S}tatistical {C}omputing
  in {P}ython}. In \bibinfo{booktitle}{\emph{{P}roceedings of the 9th {P}ython
  in {S}cience {C}onference}}, \bibfield{editor}{\bibinfo{person}{{S}t\'efan
  van~der {W}alt} {and} \bibinfo{person}{{J}arrod {M}illman}} (Eds.).
  \bibinfo{pages}{56 -- 61}.
\newblock
\urldef\tempurl%
\url{https://doi.org/10.25080/Majora-92bf1922-00a.}
\showDOI{\tempurl}


\bibitem[{Yarnold, Barbara M.}(1990)]%
        {yarnold_barbara_m_federal_1990}
\bibfield{author}{\bibinfo{person}{{Yarnold, Barbara M.}}}
  \bibinfo{year}{1990}\natexlab{}.
\newblock \showarticletitle{Federal {Court} {Outcomes} in {Asylum}-{Related}
  {Appeals} 1980-1987: {A} {Highly} ‘{Politicized}’ {Process}.”}.
\newblock \bibinfo{journal}{\emph{Springer}} \bibinfo{volume}{23},
  \bibinfo{number}{4} (\bibinfo{date}{Nov.} \bibinfo{year}{1990}),
  \bibinfo{pages}{291--306}.
\newblock
\urldef\tempurl%
\url{https://www.jstor.org/stable/pdf/4532204.pdf?refreqid=excelsior\%3A2d4deaa66899c708d6b75fa82db4997e\&ab_segments=&origin=}
\showURL{%
\tempurl}


\end{thebibliography}

\end{document}